\documentclass[13pt]{article}

\usepackage[utf8]{inputenc}
\usepackage[T1]{fontenc}
\usepackage{hyphenat}
\usepackage{xspace}
\usepackage{amsmath}
\usepackage{amsfonts}
\usepackage{hyperref}
\usepackage{url}
\usepackage{booktabs}
\usepackage{multirow}
\usepackage{subfigure}
\usepackage{makecell}
\usepackage{caption}
\usepackage{minibox}
\usepackage{bbm}
\usepackage{graphicx}
\usepackage{balance}
\usepackage{mathtools}
\usepackage{color}
\usepackage{marvosym}
\usepackage{ifthen}
\usepackage{textcomp}
\usepackage{enumitem}
\usepackage{verbatim}
\usepackage{algorithm}
\usepackage{algorithmic}
\usepackage{numprint}
\usepackage{balance}

\usepackage{amsthm}
\theoremstyle{plain}

\newcommand{\chatoDisplayMode}[1]{#1}

\definecolor{MyRed}{rgb}{0.6,0.0,0.0} 
\definecolor{MyBlack}{rgb}{0.1,0.1,0.1} 
\newcommand{\inred}[1]{{\color{MyRed}\sf\textbf{\textsc{#1}}}}
\newcommand{\frameit}[2]{
  \begin{center}
  {\color{MyRed}
  \framebox[.9\columnwidth][l]{
    \begin{minipage}{.85\columnwidth}
    \inred{#1}: {\sf\color{MyBlack}#2}
    \end{minipage}
  }\\
  }
  \end{center}
}

\newcommand{\note}[2][]{\chatoDisplayMode{\def\@tmpsig{#1}\frameit{{\Pointinghand} Note}{#2\ifx \@tmpsig \@empty \else \mbox{ --\em #1}\fi}}}
\newcommand{\todo}[2][]{\chatoDisplayMode{\def\@tmpsig{#1}\frameit{{\Writinghand} To-do}{#2\ifx \@tmpsig \@empty \else \mbox{ --\em #1}\fi}}}

\newcommand{\abbrevStyle}[1]{#1}

\newcommand{\ie}{\abbrevStyle{i.e.}\xspace}

\newcommand{\textcite}[1]{\citeauthor{#1} \shortcite{#1}}

\newcommand{\hide}[1]{}

\makeatletter
\newcommand{\iffont}[2]{\ifthenelse{\equal{\f@family}{#1}}{#2}{}}
\makeatother

\iffont{ptm}{
  \usepackage{mathptmx}

  \DeclareSymbolFont{greek}{OML}{cmm}{m}{n}
  \DeclareMathSymbol{\alpha}{\mathalpha}{greek}{"0B}
  \DeclareMathSymbol{\beta}{\mathalpha}{greek}{"0C}
  \DeclareMathSymbol{\gamma}{\mathalpha}{greek}{"0D}
  \DeclareMathSymbol{\delta}{\mathalpha}{greek}{"0E}
  \DeclareMathSymbol{\epsilon}{\mathalpha}{greek}{"0F}
  \DeclareMathSymbol{\zeta}{\mathalpha}{greek}{"10}
  \DeclareMathSymbol{\eta}{\mathalpha}{greek}{"11}
  \DeclareMathSymbol{\theta}{\mathalpha}{greek}{"12}
  \DeclareMathSymbol{\iota}{\mathalpha}{greek}{"13}
  \DeclareMathSymbol{\kappa}{\mathalpha}{greek}{"14}
  \DeclareMathSymbol{\lambda}{\mathalpha}{greek}{"15}
  \DeclareMathSymbol{\mu}{\mathalpha}{greek}{"16}
  \DeclareMathSymbol{\nu}{\mathalpha}{greek}{"17}
  \DeclareMathSymbol{\xi}{\mathalpha}{greek}{"18}
  \DeclareMathSymbol{\pi}{\mathalpha}{greek}{"19}
  \DeclareMathSymbol{\rho}{\mathalpha}{greek}{"1A}
  \DeclareMathSymbol{\sigma}{\mathalpha}{greek}{"1B}
  \DeclareMathSymbol{\tau}{\mathalpha}{greek}{"1C}
  \DeclareMathSymbol{\upsilon}{\mathalpha}{greek}{"1D}
  \DeclareMathSymbol{\phi}{\mathalpha}{greek}{"1E}
  \DeclareMathSymbol{\chi}{\mathalpha}{greek}{"1F}
  \DeclareMathSymbol{\psi}{\mathalpha}{greek}{"20}
  \DeclareMathSymbol{\omega}{\mathalpha}{greek}{"21}
  \DeclareMathSymbol{\varepsilon}{\mathalpha}{greek}{"22}
  \DeclareMathSymbol{\vartheta}{\mathalpha}{greek}{"23}
  \DeclareMathSymbol{\varpi}{\mathalpha}{greek}{"24}
  \DeclareMathSymbol{\varrho}{\mathalpha}{greek}{"25}
  \DeclareMathSymbol{\varsigma}{\mathalpha}{greek}{"26}
  \DeclareMathSymbol{\varphi}{\mathalpha}{greek}{"27}
  \DeclareSymbolFont{otone}{OT1}{cmr}{m}{n}
  \DeclareMathSymbol{\Gamma}{\mathalpha}{otone}{0}
  \DeclareMathSymbol{\Delta}{\mathalpha}{otone}{1}
  \DeclareMathSymbol{\Theta}{\mathalpha}{otone}{2}
  \DeclareMathSymbol{\Lambda}{\mathalpha}{otone}{3}
  \DeclareMathSymbol{\Xi}{\mathalpha}{otone}{4}
  \DeclareMathSymbol{\Pi}{\mathalpha}{otone}{5}
  \DeclareMathSymbol{\Sigma}{\mathalpha}{otone}{6}
  \DeclareMathSymbol{\Upsilon}{\mathalpha}{otone}{7}
  \DeclareMathSymbol{\Phi}{\mathalpha}{otone}{8}
  \DeclareMathSymbol{\Psi}{\mathalpha}{otone}{9}
  \DeclareMathSymbol{\Omega}{\mathalpha}{otone}{10}
  \DeclareSymbolFont{syms}{OML}{cmm}{m}{it}
  \DeclareMathSymbol{\partial}{\mathord}{syms}{"40}
  \DeclareMathAlphabet{\mathbold}{OML}{cmm}{b}{it}
  \DeclareSymbolFont{largesymbols}{OMX}{cmex}{m}{n}

}

\newcommand*{\crosssymbol}{%
    \text{%
      \raisebox{1ex}{%
        \makebox[0pt][l]{%
          \rule[-.2pt]{.75ex}{.4pt}%
        }%
        \makebox[.75ex]{%
          \rule[-1ex]{.4pt}{1.5ex}%
        }%
      }%
    }%
}

\usepackage{arxiv}
\usepackage[utf8]{inputenc} 
\usepackage[T1]{fontenc} 
\usepackage{hyperref} 
\usepackage{url} 
\usepackage{booktabs} 
\usepackage{amsfonts} 
\usepackage{nicefrac} 
\usepackage{microtype} 
\usepackage{lipsum}
\usepackage{graphicx}
\usepackage{caption}
\usepackage{subfigure}
\usepackage{amsmath}

\title{
Experts and authorities receive disproportionate attention on Twitter during the COVID-19 crisis
}

\author{
Kristina Gligorić,$^{1 \crosssymbol *}$
Manoel Horta Ribeiro,$^{1 \crosssymbol *}$
Martin Müller,$^{2 \crosssymbol}$ \\
\textbf{Olesia Altunina,$^{2}$ Maxime Peyrard,$^{1}$ Marcel Salathé,$^{2}$ Giovanni Colavizza,$^3$ Robert West$^1$}\\
$1:$ Data Science Lab, Ecole polytechnique fédérale de Lausanne (EPFL), Lausanne, Switzerland \\
$2:$ Digital Epidemiology Lab, Ecole polytechnique fédérale de Lausanne (EPFL), Geneva, Switzerland \\
$3:$ Institute for Logic, Language and Computation (ILLC), University of Amsterdam, Amsterdam, Netherlands \\
$\crosssymbol:$ Equal contribution \\
$*:$ Correspondence to: kristina.gligoric@epfl.ch, manoel.hortaribeiro@epfl.ch \vspace{4mm}\\
\textit{This is a pre-print. Code and reproducibility data to be made available at:}\\ \url{https://github.com/digitalepidemiologylab/experts-covid19-twitter}
\vspace{4mm}
}

\begin{document}

\maketitle

\begin{abstract}
    Timely access to accurate information is crucial during the COVID-19 pandemic. Prompted by key stakeholders’ cautioning against an “infodemic”, we study information sharing on Twitter from January through May 2020. We observe an overall surge in the volume of general as well as COVID-19-related tweets around peak lockdown in March/April 2020. With respect to engagement (retweets and likes), accounts related to healthcare, science, government and politics received by far the largest boosts, whereas accounts related to religion and sports saw a relative decrease in engagement. While the threat of an “infodemic” remains, our results show that social media also provide a platform for experts and public authorities to be widely heard during a global crisis.
\end{abstract}

The effective communication of trustworthy information has proven key to overcome public health crises in the past, especially when the coordinated effort of entire populations was required (1). At the same time, the widespread adoption of social media has been linked to the spread of low-quality, mis-, and disinformation (2, 3), with some studies concluding that fake information goes viral more easily, and has broader reach, than trustworthy information on social media (4,5). These findings are particularly pertinent to the COVID-19 crisis, which is unfolding during a time of unprecedented Internet penetration and has drawn enormous attention on both traditional and social media (6,7,8), a fact that has led the World Health Organization to declare a state of “infodemic” (9), stating that “people must have access to accurate information to protect themselves and others” (10).
 
When considering global health crises, experts and public institutions are considered to be trusted information sources (11). Their recognition might, however, be diminished by the spread of low-quality or false information in social media (12). In a crisis as unprecedented as the COVID-19 outbreak, the prevalence of trusted information sources on social media may play a critical role in shaping an effective response. Within the space of online social media, Twitter plays a key role (13) and has proven to be effective for monitoring ongoing crises (12), including pandemics (14-16). Preliminary work on Twitter during the COVID-19 pandemic in fact suggests that, while false information is tweeted more than science-based evidence, the latter is shared more via retweets (17).

The overall goal of the present research is to map the Twitter landscape during the COVID-19 pandemic from an account-centric angle: who speaks and who is being heard? To better understand the plurality of voices taking part in the public debate on Twitter, we developed a custom taxonomy of categories of user accounts (see legend of Fig. 1A for a list of categories; see SI 1.2.2 for details on how the taxonomy was derived). We employed Twitter’s complete COVID-19 streaming endpoint, to which access was granted starting 6 May 2020. The stream includes all tweets containing one of 590 multilingual keywords related to COVID-19. The population we study consists of all user accounts that posted COVID-19-related content that has received a non-negligible amount of attention. Fig. 1A summarizes the specific study design that was implemented. Based on the first full week of the COVID-19 stream (6–12 May 2020; “account sampling period” in Fig. 1A), we constructed a sample of 14,200 Twitter accounts that each had posted at least one COVID-19-related tweet with at least 10 retweets or likes (henceforth, “engagements”), and annotated each account in the sample with its category using crowdsourcing (SI 1.2.3). The sample was constructed to be representative of the overall population (SI 1.2.1). The distribution over account categories is plotted in Fig. 1B. We then queried Twitter’s application programming interface (API) to collect all tweets—regardless of whether they contained a COVID-19 keyword—for the 14,200 sampled accounts during the 5-month period from 1 January to 31 May 2020. The first 2 weeks were used as a “baseline period” to calibrate accounts’ behavior, which was tracked during the following 4.5 months (“study period” in Fig. 1A).

\begin{figure}[t]
\centering
\begin{minipage}{0.71\textwidth}
  \centering
  \includegraphics[width=\linewidth]{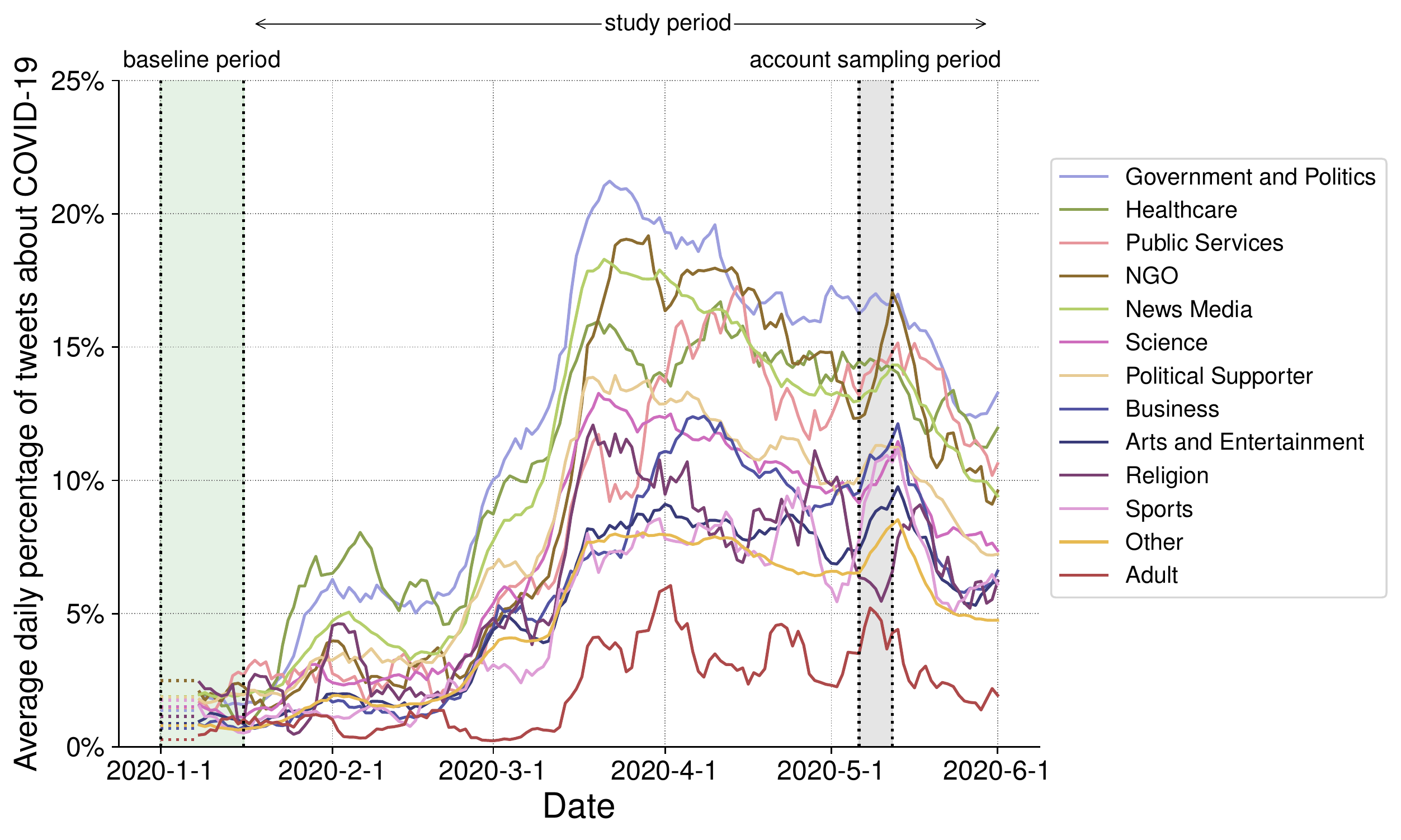}
  \caption*{(A)}
\end{minipage}%
\begin{minipage}{0.29\textwidth}
  \centering
  \includegraphics[width=\linewidth]{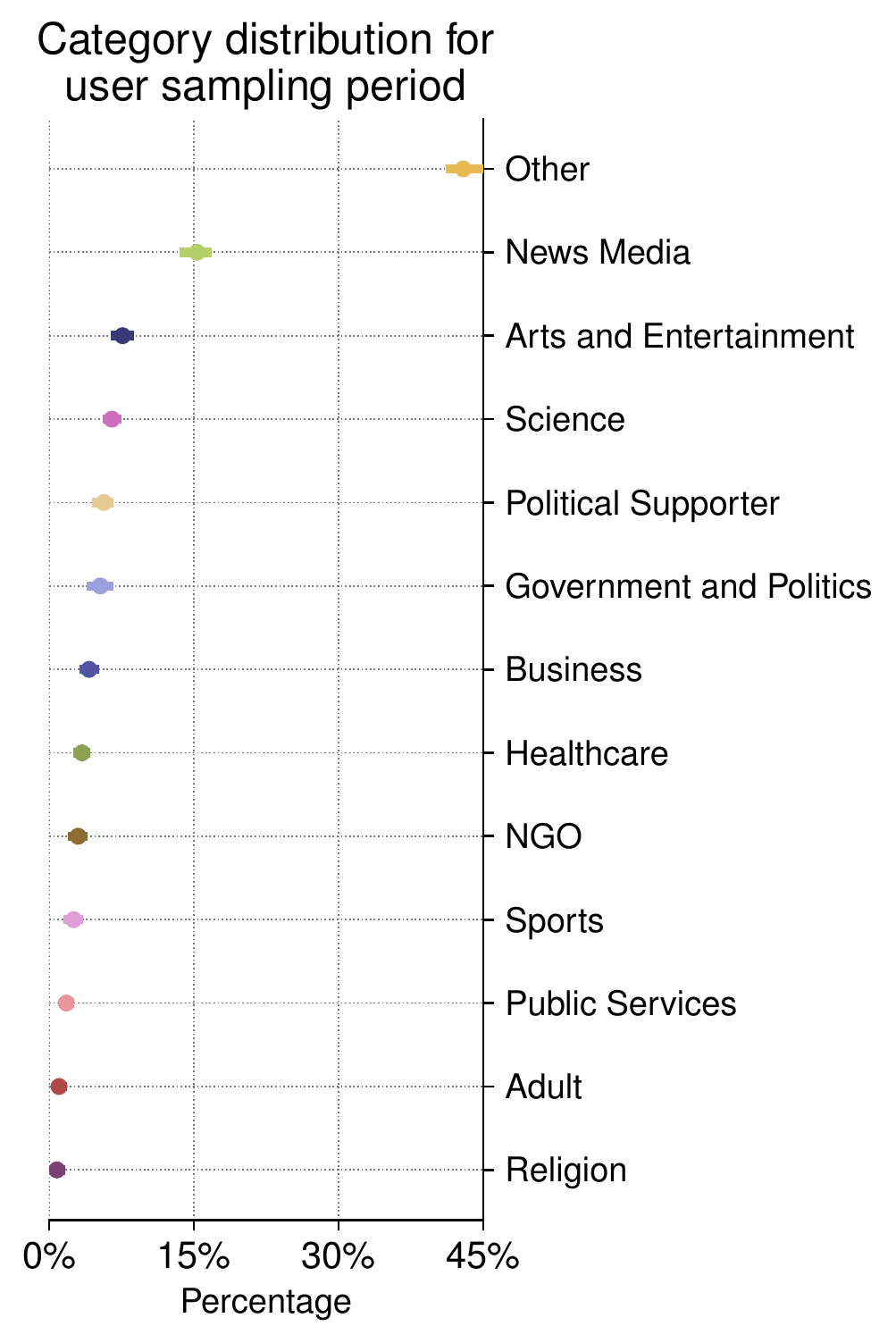}
  \caption*{(B)}
\end{minipage}
\caption{\textbf{Study design.} (A) We study Twitter accounts that posted at least one COVID-19-related tweet that received at least 10 retweets + likes during the week of 6–12 May 2020 (account sampling period, shaded gray). We create a sample of these accounts, categorize them into 13 categories (cf. legend), and collect their entire Twitter timelines from 1 January to 31 May 2020. The first 14 days serve as a baseline period (shaded green), and the remaining 4.5 months, as the study period. Inverse probability weighting (see SI 1.3) is used throughout all analyses to make the sample representative. Lines in (A) represent the percentage of tweets related to COVID-19 per category for the sampled accounts (7-day moving averages; giving every account equal weight). Starting March 2020, a substantial fraction of tweets refers to the pandemic. (B) Distribution over categories among all accounts in the account sampling period, estimated from the manually categorized, representative sample.}
\end{figure}

First, in Fig. 1A, we track the fraction of tweets containing a COVID-19 keyword, macro-averaged over all accounts per category. Whereas all categories posted very small fractions (0\%–2.5\%) of COVID-19-related content during the baseline period in early January 2020, the topic became more prevalent in late January and peaked between mid March and early April, when up to 1 in 5 tweets contained a COVID-19 keyword for some categories, with the highest peaks observed for categories of direct relevance for the pandemic: Government \& Politics (peak 21\%), NGOs (19\%), News Media (18\%), Public Services (17\%), and Healthcare (17\%). Less directly relevant categories also referred to COVID-19 in considerable fractions of their posts, e.g., Religion (12\%), Sports (11\%), and Arts \& Entertainment (10\%). This first result highlights the deep impact the COVID-19 pandemic has had on the Twitter ecosystem.

Next, we investigate whether the studied accounts have changed their overall tweeting frequency during the COVID-19 pandemic. This analysis considers all tweets posted by the studied accounts, regardless of whether they contain a COVID-19 keyword or not. We calibrated an account’s tweet volume during the baseline period and computed, for each subsequent week, the percentage change over the baseline (SI 1.1). The results, visualized as blue curves in Fig. 2, show that tweet volume increased considerably for all categories, compared to the pre-pandemic baseline. The most notable cases are Religion, which peaked at +207\%, and Healthcare, at +175\%. Even the least affected categories showed a strong increase, with News Media peaking at +63\%, and Arts \& Entertainment, at +73\%.

\begin{figure}
    \centering
    \includegraphics[width=\textwidth]{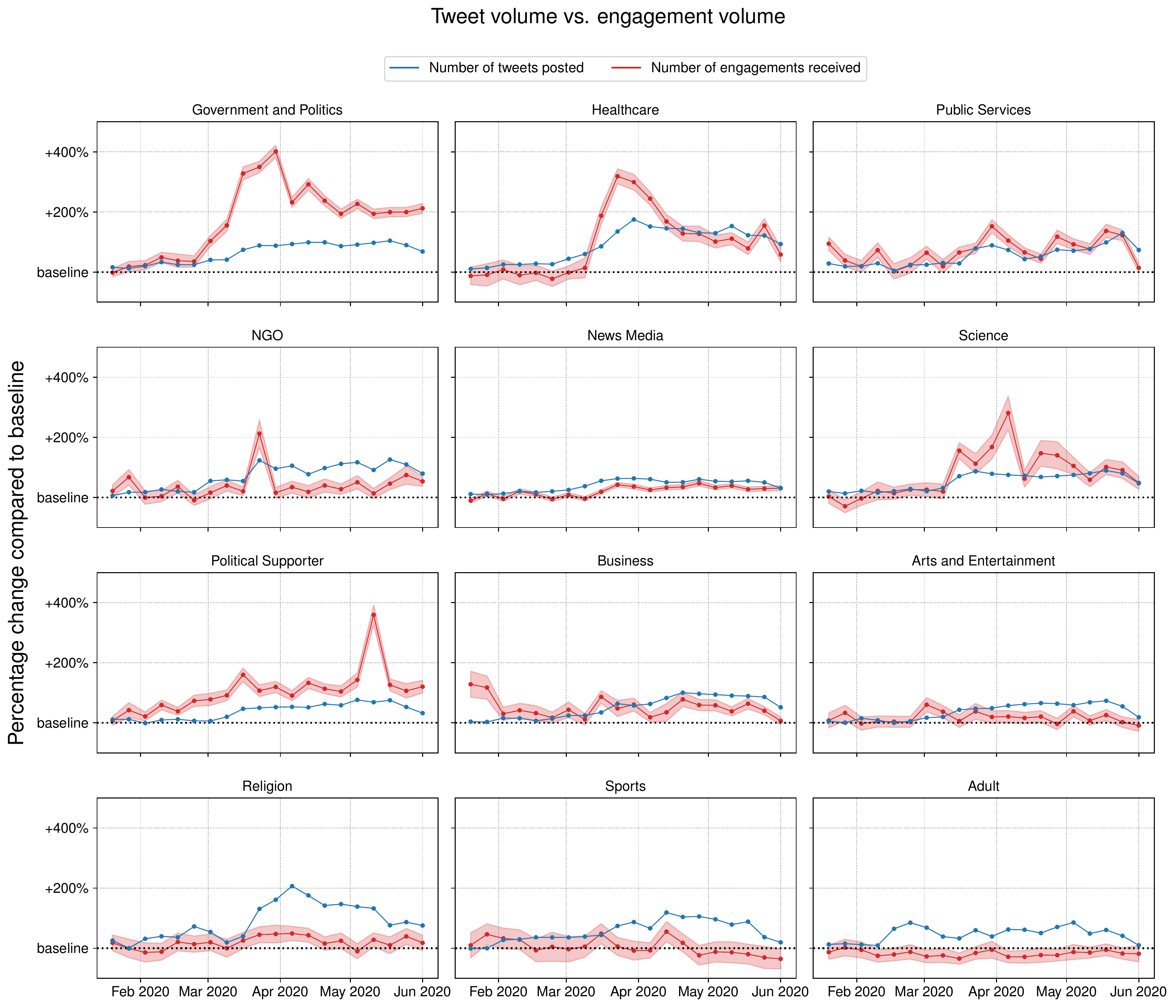}
    \caption{\textbf{Tweet volume vs. engagement volume.} Weekly percentage increase over the early-January baseline (cf. Fig. 1A) with respect to the number of tweets posted (blue) and engagements (retweets + likes) received (red) (macro-averages over accounts; estimated from the representative sample of accounts, cf. Fig. 1A; with 95\% confidence intervals). In all categories, tweet volumes (blue) rise far above baseline, particularly starting mid March 2020, when lockdowns are imposed worldwide. Engagement (red) behaves more heterogeneously, rising more for categories of particular structural importance to the pandemic (related to health, science, government, politics). Red lying above [below] blue corresponds to a rate of engagements-per-tweet that is higher [lower] than at baseline. That is, Government \& Politics and Political Supporters see lasting, whereas Health and Science see transient, boosts in engagements-per-tweet.}
\end{figure}

In order to determine to what extent the increase in tweeting is associated with an increase in being noticed, we also measured the engagement (retweets + likes) received by each account, again calibrated against the pre-pandemic baseline. The results, visualized as red curves in Fig. 2, show that engagement volume behaved more heterogeneously than tweet volume. Some categories saw substantial increases in engagement—much larger than the respective increases in tweet volume (red above blue in Fig. 2). In particular, for Government \& Politics, the increase in engagement peaked at +402\%, whereas the increase in tweet volume peaked at only +102\%. Similar effects were observed for Healthcare (+319\% vs. +175\%), Science (+281\% vs. +89\%), and Political Supporters (+359\% vs. +76\%). Accounts in these categories thus became, on average, more “effective” at tweeting, with a higher number of engagements per tweet than at baseline. Conversely, for other categories, engagement per tweet decreased with the pandemic (red below blue in Fig. 2). Most notably, Religion saw little increase in engagement (peak +49\%), despite having increased its tweet volume most out of all the categories (peak +207\%). Similar effects were observed for Sports (+55\% vs. 119\%) and Adult content (+2\% vs. +86\%). It is noteworthy that, among the “effective” categories, two distinct patterns emerge: on the one hand, for Healthcare and Science, the blue curve in Fig. 2 converges to nearly the same value as the red curve (i.e., engagement per tweet reverts to the level of the baseline period), whereas, on the other hand, for Government \& Politics and Political Supporters, the red curve remains consistently above the blue curve (i.e., engagement per tweet stays above the level of the baseline period). To summarize, Healthcare and Science saw transient, whilst Government \& Politics and Political Supporters saw persistent, boosts in engagement that far exceeded the respective boosts in tweet volumes. On the contrary, Religion, Sports, and Adult content accounts saw a decrease in engagement, despite the fact that they, too, tweeted more.

To directly compare categories to each other, we computed two global rankings of accounts (both computed 1–4 June 2020, when account timelines were collected), one with respect to engagement counts, the other with respect to follower counts. Average ranks (normalized such that 1 and 0 correspond to top and bottom, respectively) are plotted for all categories in Fig. 3. We will discuss the follower-count ranking (x-axis) later, and for now focus on the engagement ranking (y-axis). Average engagement ranks were significantly (p < 0.05, two-sided KS tests) higher for tweets from the study period (end points of arrows) than for tweets from the baseline period (starting points of arrows) for Healthcare, Science, Government \& Politics, Political Supporters, Public Services, and News Media, whereas the effect was reversed for Religion, Sports, Adult content, and Business. While these results echo the findings from Fig. 2, they also add nuance: as all accounts participated in the rank computations, Fig. 3 may be considered a “zero-sum game”, in the sense that one account’s increase must be offset by another account’s decrease. Viewed in this light, Fig. 3 suggests that Healthcare, Science, Government \& Politics, etc., have gained attention relative to Religion, Sports, and Adult content.

Follower counts on Twitter vary widely across accounts (18). The intuitive expectation that a larger follower count is associated with more engagement is overall confirmed by Fig. 3, with a category-level Spearman rank correlation of 0.71 (p = 0.0067, t(11) = 3.33) in the baseline period, and 0.62 (p = 0.024, t(11) = 2.62) in the study period. Some important exceptions, however, emerge: Healthcare accounts on average rank lowest with respect to follower count during the study period (12 out of 12 when ignoring the “Other” category), but rank in the upper half (6 out of 12) with respect to engagement. The opposite effect is observed for Sports, Arts \& Entertainment, and Adult content, which are in the top half with respect to follower count, but in the bottom half with respect to engagement. These findings suggest that the increased attention to categories that are most directly important in the fight against the pandemic is not merely a consequence of the size of their follower base.

In conclusion, our work shows that Twitter accounts associated with structurally important roles are “boosted” during the pandemic. While accounts in all categories on average increased their tweet volume, accounts related to Science, Healthcare, and Government \& Politics saw the largest boosts in engagement. Despite the overall surge in produced content, which could be symptomatic of an infodemic, our findings imply that users selectively promote information from structurally relevant sources during the crisis. The ways in which accounts from different structurally important categories are boosted seem to differ. As the crisis is shifting from a health crisis to a societal crisis, accounts related to healthcare and science receive progressively less attention, whereas attention to governments and politicians remains high. A caveat to our analysis is that it is based on self-declared account descriptions on Twitter and does not take into account the content of messages. This is an important direction for future work, since within categories, the quality of specific messages and the alignment with the scientific consensus could vary. Ultimately, while concerns about the spread of misinformation and the associated fears of an “infodemic” deserve our continued attention, we show that, to date, Twitter users have disproportionately paid attention to experts and authorities in the COVID-19 crisis.

\begin{figure}
    \centering
    \includegraphics[width=\textwidth]{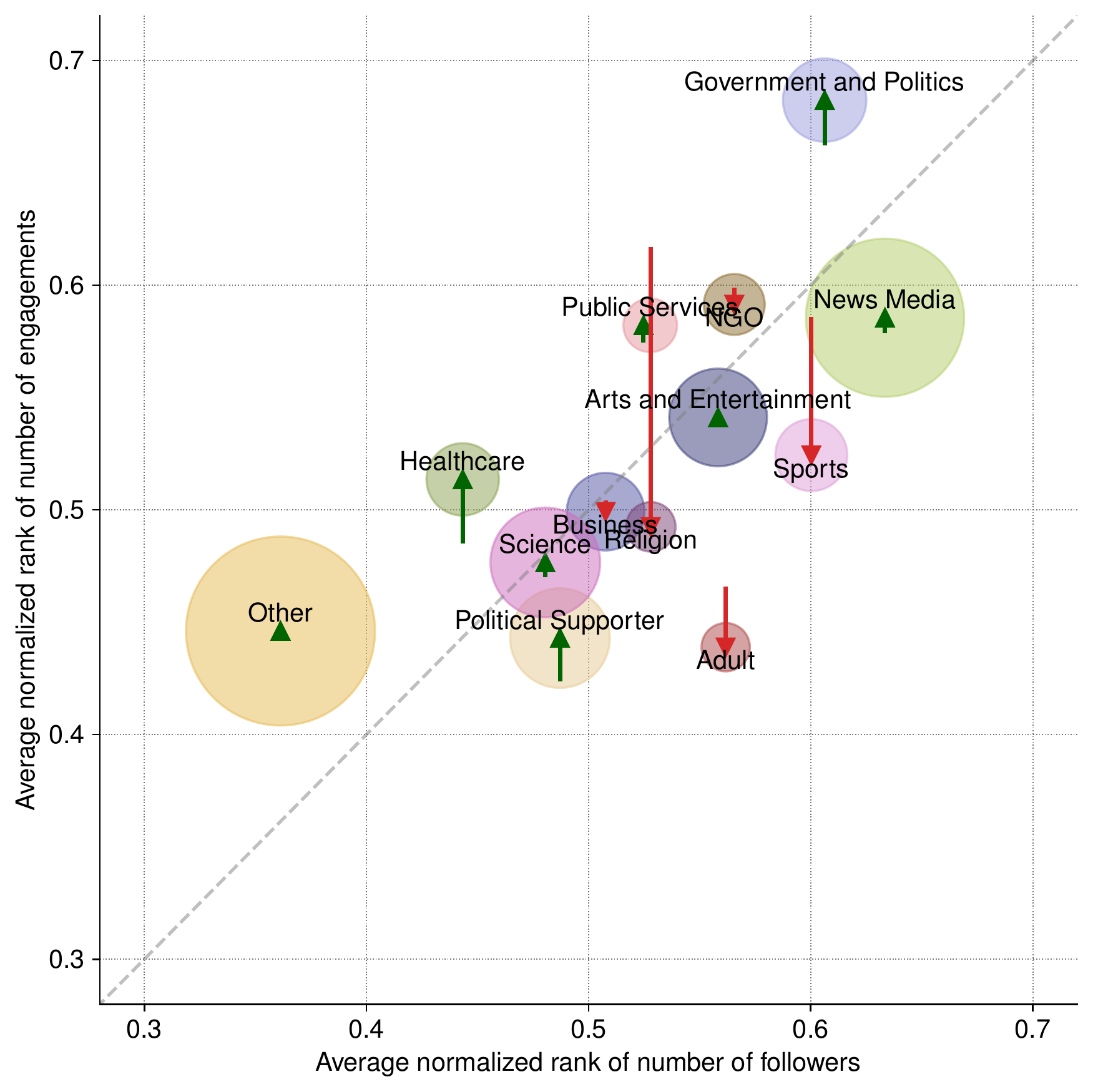}
    \caption{\textbf{Rank-based comparison of account categories.} The y-axis shows normalized ranks with respect to the number of engagements (retweets + likes) received for tweets posted during the baseline period (arrow starting points) and for tweets posted during the study period (arrow end points), averaged over the accounts in the respective category. The x-axis shows ranks with respect to follower counts (as observed after the end of the study period, 1–4 June 2020). Ranks were normalized such that 1 and 0 correspond to top and bottom, respectively. Disk radius is proportional to the number of tweets posted by the category in the study period. Categories of particular structural importance to the pandemic (related to health, science, government, politics, news) have risen (upward arrows), whereas Religion, Sports, and Adult have fallen (downward arrows). Healthcare, Government \& Politics, Public Services, and NGOs are particularly much engaged-with, relative to their follower counts (position above diagonal).}
\end{figure}

\noindent
\textbf{REFERENCES}

\begin{enumerate}
    \item H. W. Park, S. Park, M. Chong. Conversations and medical news frames on twitter: Infodemiological study on Covid-19 in South Korea. Journal of Medical Internet Research 22, e18897 (2020).
\item D. M. J. Lazer, M. A. Baum, Y. Benkler, A. J. Berinsky, K. M. Greenhill, F. Menczer, M. J. Metzger, B. Nyhan, G. Pennycook, D. Rothschild, M. Schudson, S. A. Sloman, C. R. Sunstein, E. A. Thorson, D. J. Watts, J. L. Zittrain. The science of fake news. Science 359, 1094-1096 (2018).
\item B. Swire-Thompson, D. Lazer. Public health and online misinformation: challenges and recommendations. Annual Review of Public Health, 41, 433-451 (2020).
\item S. Vosoughi, D. Roy, S. Aral. The spread of true and false news online. Science 359, 1146-1151 (2018).
\item M. Del Vicario, A. Bessi, F. Zollo, F. Petroni, A. Scala, G. Caldarelli, H. E. Stanley, W. Quattrociocchi. The spreading of misinformation online. Proceedings of the National Academy of Science 113, 554-559 (2016).
\item A. Depoux, S. Martin, E. Karafillakis, R. Preet, A. Wilder-Smith, H. Larson. The pandemic of social media panic travels faster than the COVID-19 outbreak. Journal of Travel Medicine 27, 10.1093 (2020).
\item E. Chen, K. Lerman, E. Ferrara. Tracking Social Media Discourse About the COVID-19 Pandemic: Development of a Public Coronavirus Twitter Data Set. JMIR Public Health and Surveillance 6. (2020).
\item K.C Yang, C. Torres-Lugo, F. Menczer. Prevalence of Low-credibility Information on Twitter during the Covid-19 Outbreak. Proceedings of ICWSM International Workshop on Cyber Social Threats. (2020)
\item World Health Organization. 2020. Novel Coronavirus (2019-nCoV). Situation Report 13, 2 February. www.who.int/docs/default-source/coronaviruse/situation-reports/20200202-sitrep-13-ncov-v3.pdf
\item World Health Organization. 2020. Director-General’s remarks at the media briefing on 2019 novel coronavirus on 8 February 2020. https://www.who.int/dg/speeches/detail/director-general-s-remarks-at-the-media-briefing-on-2019-novel-coronavirus---8-february-2020 
\item M. Dutta-Bergman. Trusted online sources of health information: Differences in demographics, health beliefs, and health-information orientation. Journal of Medical Internet Research 5, 3 (2003).
\item C. Reuter, M. Kaufhold. Fifteen years of social media in emergencies: A retrospective review and future directions for crisis informatics. Journal of Contingencies and Crisis Management 26, 41-57 (2018).
\item D. A. Broniatowski, M. J. Paul, M. Dredze. Twitter: Big data opportunities. Science 345, 148-148 (2014).
\item M. Wagner, V. Lampos, I. J. Cox, R. Pebody. The added value of online user-generated content in traditional methods for influenza surveillance. Scientific Reports 8, 13963 (2018).
\item S. Shin, D. Seo, J. An, H. Kwak, S. Kim, J. Gwack, M. Jo. High correlation of Middle East respiratory syndrome spread with Google search and Twitter trends in Korea. Scientific Reports 6, 32920 (2016).
\item M. Salathé, C. C. Freifeld, S. R. Mekaru, A. F. Tomasulo, J. S. Brownstein. Influenza A (H7N9) and the importance of digital epidemiology. The New England Journal of Medicine 369, 401-404 (2013).
\item C. M. Pulido, B. Villarejo-Carballido, G. Redondo-Sama, A. Gómez. COVID-19 Infodemic: More retweets for science-based information on coronavirus than for false information. International Sociology 35, 377-392 (2020).
\item  M. Cha, H. Haddadi, F. Benevenuto, K. P. Gummadi. Measuring User Influence in Twitter: The Million Follower Fallacy. AAAI Conference on Weblogs and Social Media (ICWSM) 4, 10-17 (2010).

\end{enumerate}

\clearpage
\captionsetup[table]{name=Table S}
\captionsetup[figure]{name=Fig. S}
\section{Materials and Methods}
\label{sec:methods}
\subsection{Description of the data}

\begin{figure}[h]
    \centering
    \includegraphics[width = \textwidth]{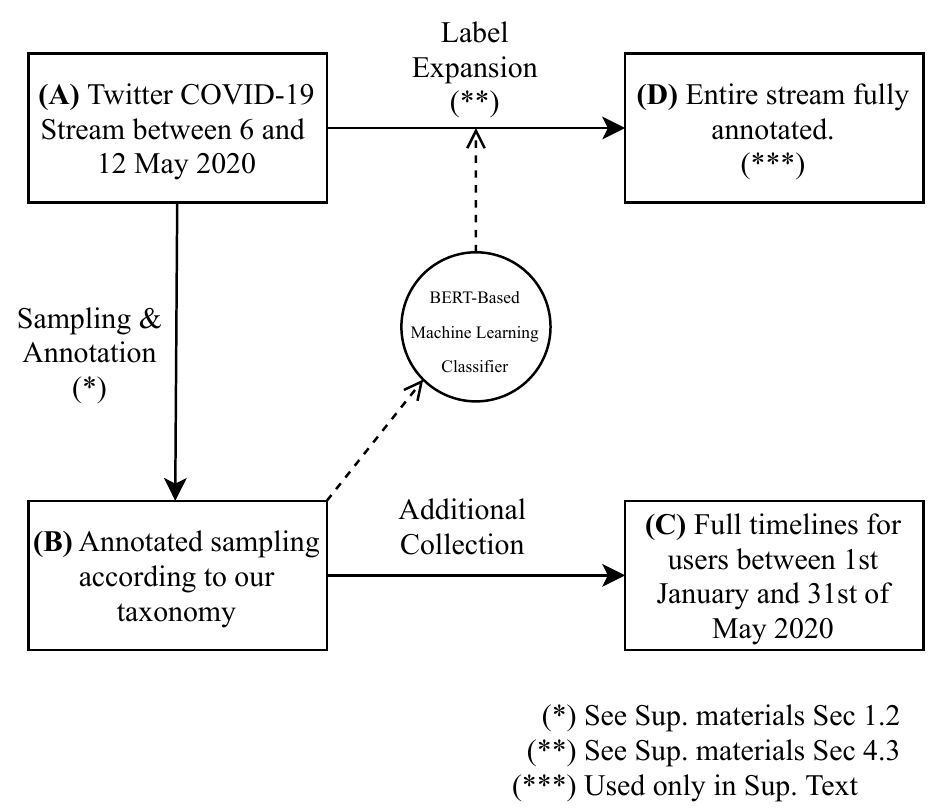}
    
    \caption{Diagram with the original and derived datasets we used. 
    \textbf{(A)} Our original data source is comprised of all tweets from the COVID-19 stream between the 6th and the 12th of May 2020.
    \textbf{(B)} We then sample a fraction of these accounts and annotate them according to the taxonomy we developed.
    \textbf{(C)} For the annotated accounts, we additionally collect their entire timelines between the 1st of January to the 31st of May 2020.
    \textbf{(D)} Lastly, we leverage the annotated sample to train a machine learning classifier which is used to classify the remaining accounts in the entire week for the COVID-19 stream.
    }
    \label{fig:example}
\end{figure}

Fig. S~\ref{fig:example} presents a diagram with all the original and derived data sources used. We employ Twitter's complete COVID-19 streaming endpoint,%
\footnote{ Announced at: {\scriptsize \url{https://blog.twitter.com/developer/en_us/topics/tools/2020/covid19_public_conversation_data.html}}} which was made available to researchers upon request (Dataset \textbf{A}). 
The endpoint includes all tweets containing one of several multilingual keywords -- curated by Twitter -- related to COVID-19, as well as all retweets and replies to those tweets.%
\footnote{The complete list of keywords is available at: {\scriptsize \url{ https://developer.twitter.com/en/docs/labs/covid19-stream/overview}}}

We focus on COVID-19 tweets posted during the week from May 6th to May 12th, 2020, written in ten major languages: English, Japanese, Spanish, Portuguese, French, German, Italian, Arabic, Indonesian and Hindi.
The language of a tweet is detected by Twitter and obtained directly from the tweets object.
We limit our analysis to accounts that tweeted at least one popular COVID tweet during the week of sampling (a tweet that received at least 10 retweets).
We then perform sampling and annotation according to our taxonomy, to produce Dataset \textbf{B}: the annotated sample.

Next, we get the timelines for all such accounts in the collected sample, collecting all the tweets they posted in 2020, and we study all of their tweets posted between Jan 1st and May 31st, 2020 (Dataset \textbf{C}). To do so, we employ Twitter's API%
\footnote{\scriptsize\url{https://developer.twitter.com/en/docs/tweets/timelines/overview}} (for accounts with less than 3200 tweets between 01/01/2020 and 31/05/2020), and Twint\footnote{\scriptsize \url{https://github.com/twintproject/twint}}, a crawler that uses a Web UI for scraping (for accounts with more than 3200 in this time frame).

Additionally, we leverage the annotated sample to train a machine learning classifier which is used to expand the labels by classifying the remaining accounts in the entire week for the COVID-19 stream, to produce Dataset \textbf{D}.

Overall, we start our analysis from 467.36k tweets that received at least 10 retweets, posted during the seven account sampling period days in May, by 196.95k unique accounts (Dataset A). After sampling (Dataset B) and enriching the timelines, our the dataset C consists of 11.47M tweets (736.73k out which contain a COVID-19 keyword, using the list of COVID-19 keywords curated by Twitter).

In our analyses, we calibrate an account’s tweet volume and engagement during the baseline period and compute, for each subsequent week, the percentage change over the baseline. To account for the possibility that some days of the week (Monday, Tuesday, etc.) might generally see higher tweet volumes, calibration is done by the day of the week, for the day level analyses.

\subsection{Annotation methodology}

To better understand the attention patterns on Twitter amidst the COVID-19 crisis, we develop a taxonomy of account categories and then proceed to annotate tens of thousands of accounts using Amazon Mechanical Turk. 
We devise our taxonomy based on techniques from grounded theory, building a robust categorization scheme of Twitter accounts who participate in COVID-19 discussions.
In what follows, we describe:
1) how the account sampling was done (subsection~\ref{sec:account_sampling}); 
2) how the taxonomy was developed (subsection~\ref{sec:taxonomy}); 
3) how we annotated 14,200 accounts using crowdsourcing (subsection ~\ref{sec:crowdsourced_annotation}).

\subsubsection{Account sampling}
\label{sec:account_sampling}

For both iterative development of taxonomy and crowdsourced annotation, we first select a subsample of the accounts who posted at least one 10-retweets or more tweet about COVID-19 between the 6th and the 12th of May and who tweeted in one of the 10 most popular languages in the sample: English, Japanese, Spanish, Portuguese, Italian, Arabic, German and French, Hindi and Indonesian (Table S~\ref{tab:dist}).

\begin{enumerate}
    \item First, we restrict ourselves to studying only those accounts which posted at least one popular tweet in the 7 days. A tweet is popular if it has received at least 10 retweets. This requirement ensures that sampled accounts received a non-negligible amount of attention. Such accounts comprise 1.96\% of all accounts, 1.73\% of all tweets, and 84.05\% of all retweets, in the COVID-19 stream during the account sampling period.
    
    \item  Second, for each language (Table S~\ref{tab:dist}), we calculate quintiles for the number of followers and number of retweets. 
    By doing so, for each language, we have split accounts into 25 ``buckets'' where each bucket corresponds to a different combination of quintiles for the number of followers and of retweets.
    
    \item Third, we sampled the same number of accounts from each bucket. 
    We sample accounts across languages proportional to the log of the number of tweets in that language, so that accounts tweeting in bigger languages are not over-represented. 

    \item Lastly, we translated all account metadata from accounts that were not tweeting in English into English using Google's translation API.

\end{enumerate}

Overall, tweets that got at least 10 retweets obtain 84.05\% of all retweets on COVID tweets, so in this way, we capture the majority of the engagement COVID tweets receive in total.
 
\subsubsection{Iterative development of the taxonomy}
\label{sec:taxonomy}

Next, we explain the steps taken to develop the taxonomy.
 
\begin{itemize}
\item[\textbf{Stage 1}:] \textbf{Building the initial taxonomy.}
Before inspecting the data, the authors discussed broad relevant categories of individuals and entities likely to play a significant role in the COVID-19 online debate.
It was determined that categories have to either represent concrete occupations (researcher, medical doctor, and similar) disparately affected or in other ways essential in the context of the pandemic; or, groups of individuals or institutions that shape public discourse. 
Also, categories had to be significantly represented in the data. However, this was only considered at the end of each iteration, when considering which labels to incorporate to the taxonomy.

\item[\textbf{Stage 2}:] \textbf{Initial inspection.}
Three researchers (all authors of the paper) independently explored three different random samples of account descriptions in English, consisting of a hundred accounts each. 
This was done to build a common understanding of the type of descriptions prevalent in the data. 
We defer explaining how the samples were generated to Subsection 1.2.3.
For each account, researchers assessed the information about how the account presents itself: the description of the account, Twitter handle, and name. 
Researchers carefully analyzed the account descriptions considering the categories and wrote notes about the applicability of categories. After that, researchers shared their observations, discussed the initial categories, and adapted them.

\item[\textbf{Stage 3}:] \textbf{Iterative Coding.}
Iterative coding was done as follows. In each iteration, three researchers (all authors of this paper) annotated the same set of 100 accounts, with the possibility of expanding the category set.
Each account was to be assigned any number of categories, which were determined based on accounts' self-declaration on Twitter (we did not inspect any other information beyond the description, the account name, and the screen name).
At the end of each round, researchers individually discussed all disagreements and the overall appropriateness of the categories. 
Then, they made changes to the categories when necessary, adding new categories or tweaking the definitions of existing categories. 
Before starting the iterative coding, the researchers agreed on the criteria for stopping the iterations. All of the following three criteria had to be satisfied: 
1) Average pairwise Fleiss Kappa agreement is greater than 0.6;
2) Researchers agree that the categories are not ambiguous;
3) The difference in the prevalence of \textit{Other} between two subsequent iterations is smaller than $5\%$. 
We repeated this annotation process three times before satisfying all three criteria, the rounds yielded inter-annotator agreements of $0.6$, $0.65$, and $0.67$, respectively.
We depict the final taxonomy in Tables S~\ref{tab:cat} and S~\ref{tab:type}.
Notice that during the analyses in the paper, we collapsed some of the categories together as some were rather sparse, and as their joint interpretation was useful.
\end{itemize}

Orthogonal to categories, annotators were also asked to identify for each account, whether the account belonged to an individual or an institution. 
For this annotation, in the iterative coding stage, inter-annotator agreement scores were of $0.63$, $0.89$, and $0.83$, respectively.

\subsubsection{Crowdsourced annotation}
\label{sec:crowdsourced_annotation}

\begin{figure}
    \centering
    \includegraphics[width = \textwidth]{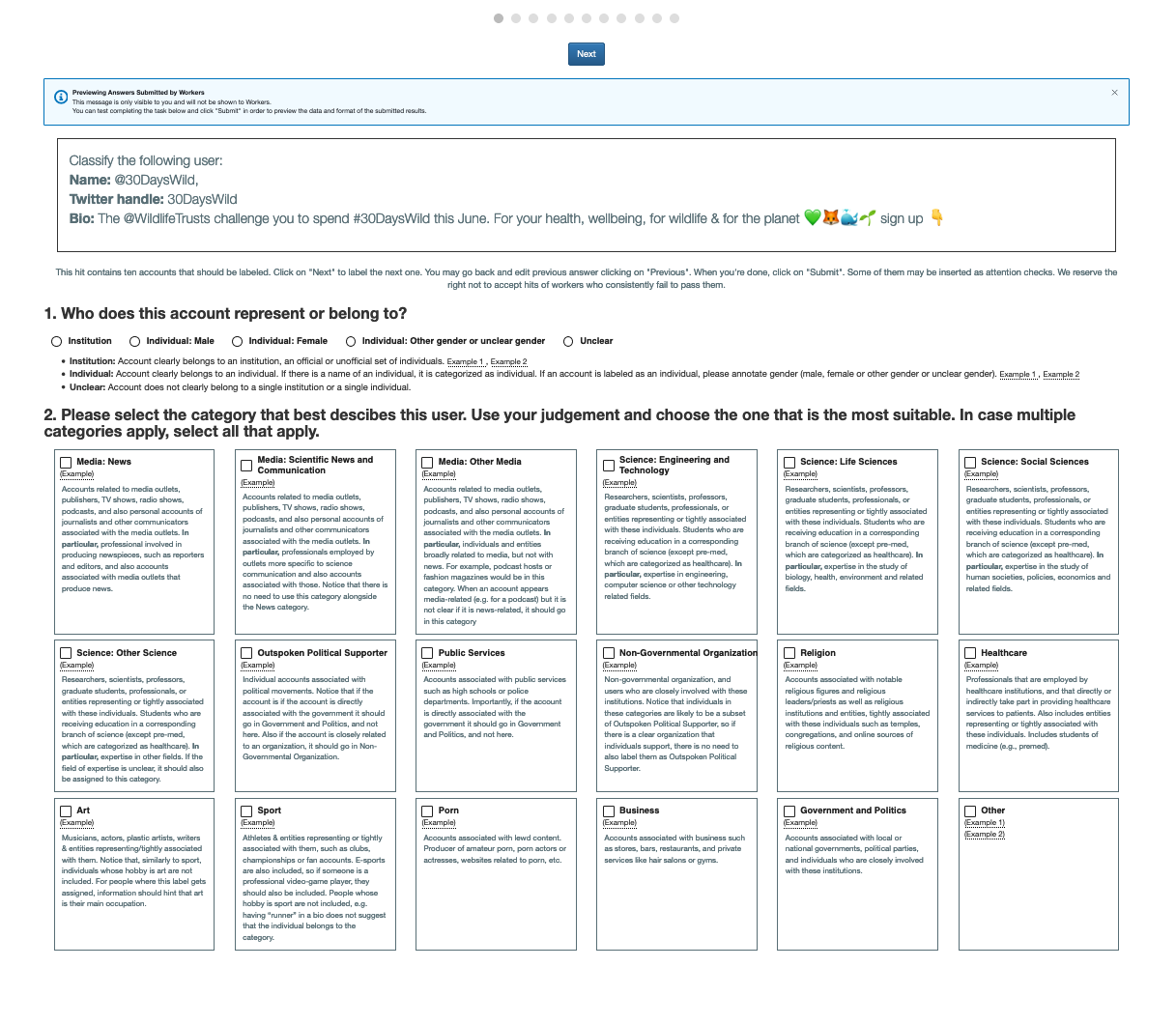}
    \caption{Screenshot of the annotation interface.}
    \label{fig:screenshot}
\end{figure}

We detail the crowdsourcing annotation process, where we annotate 14,200 sampled accounts. This amounts to 7.2\% of all accounts adhering to our restrictions, a total of 14,200 accounts out of 196,948. The human intelligence task (HIT) design is shown in Figure S~\ref{fig:screenshot}. 
Crowdsourced workers were paid 0.50 USD per HIT, and each HIT consisted of a batch of 10 different account annotation tasks. 
According to our estimates, it took 2-3 minutes to complete a single HIT, which made the compensation for the task substantially above US federal minimum wage of 7.25 \$/h.
Annotators had to select the categories from small boxes, each of which contained a description of the category, as well as a couple of explained examples (an account bio, and the reason why it would fit in a given category).

To study the feasibility of the annotation through crowdsourcing we ran a pilot where crowd-workers had to annotate the same tasks as the researchers did in their last iteration (when the categories were already set). 
We found that the results were satisfactory, majority vote label of crowd workers agreeing with the majority vote category of researchers 82\% of the time.
For the type of account (individual vs. institution) the agreement was of 91\%. Once the feasibility was established, we proceeded to annotate the accounts collecting 3 independent annotations per account.  
For accounts for whom there was no clear agreement on the category (\ie, there is no single most frequent annotation of type or category attributed by multiple workers), we collected annotation by an additional fourth annotator.
In total, we annotate 14,200 accounts belonging to 10 languages.
We report the inter-annotator agreement for each language in Table S~\ref{tab:aggr}.

For an account, we determine its dominant category as the most frequent annotation marked by at least two workers. 
If there are multiple most frequent annotations assigned by multiple workers, we break the tie randomly to choose one dominant (4.65\% of accounts). 
If there is no agreement, \ie, there is no most frequent category annotation given by at least two workers, we don't assign a dominant category annotation (7.26\% of accounts).

Finally, we limit our analysis to accounts tweeting in English, Japanese, Spanish, Portuguese, Italian, Arabic, German and French, and discard Hindi and Indonesian, as we spotted lower inter-annotator agreement compared to the other languages (less than 0.2), likely due to poorer automated translation quality.

\subsection{Inverse Probability Weighting}

In all the conducted analysis, we had to extrapolate the distribution of categories we observed in the sampled data to all the accounts. 
Recall that we divided all the tweets into 25 buckets and sampled, for each language, the same amount of accounts for each bucket. However, the buckets did not have the same amount of accounts each, and thus it may be that we over-represented some of the buckets and under-represented others. 

To address this issue, we perform an Inverse Probability Weighting scheme where we calculate the probability of being sampled, $ps$, at each bucket $k$ as:
\begin{equation}
        ps_k = \frac{\#sampled_k}{\#accounts_k}
\end{equation}

and use the inverse value, that is $ps_k^{-1}$ as the weight for all accounts in that bucket. Intuitively, this means that if we proportionally sampled twice from one of the buckets, these accounts will receive half the weight.
    
 Let $1_{\{cat,acc\}}$ be an indicator variable that indicates, for a given account and given category, whether most annotators thought the account belonged to the category. 
 To calculate the probability of a given category for a given language, we simply calculate, for all accounts of that language, the average of the indicator variable $1_{\{cat,acc\}}$ weighted according to the bucket the account was in.
    
To obtain a confidence interval, we bootstrap this calculation 1000 times. 
That is, we generate a random sample for each language obtaining $k$ accounts from each bucket (thus simulating the original sampling procedure) and then calculate the category distribution. 
We repeat it 1000 times to obtain  95\% confidence intervals. 

\begin{figure}[h]
    \centering
    \includegraphics[width=\linewidth]{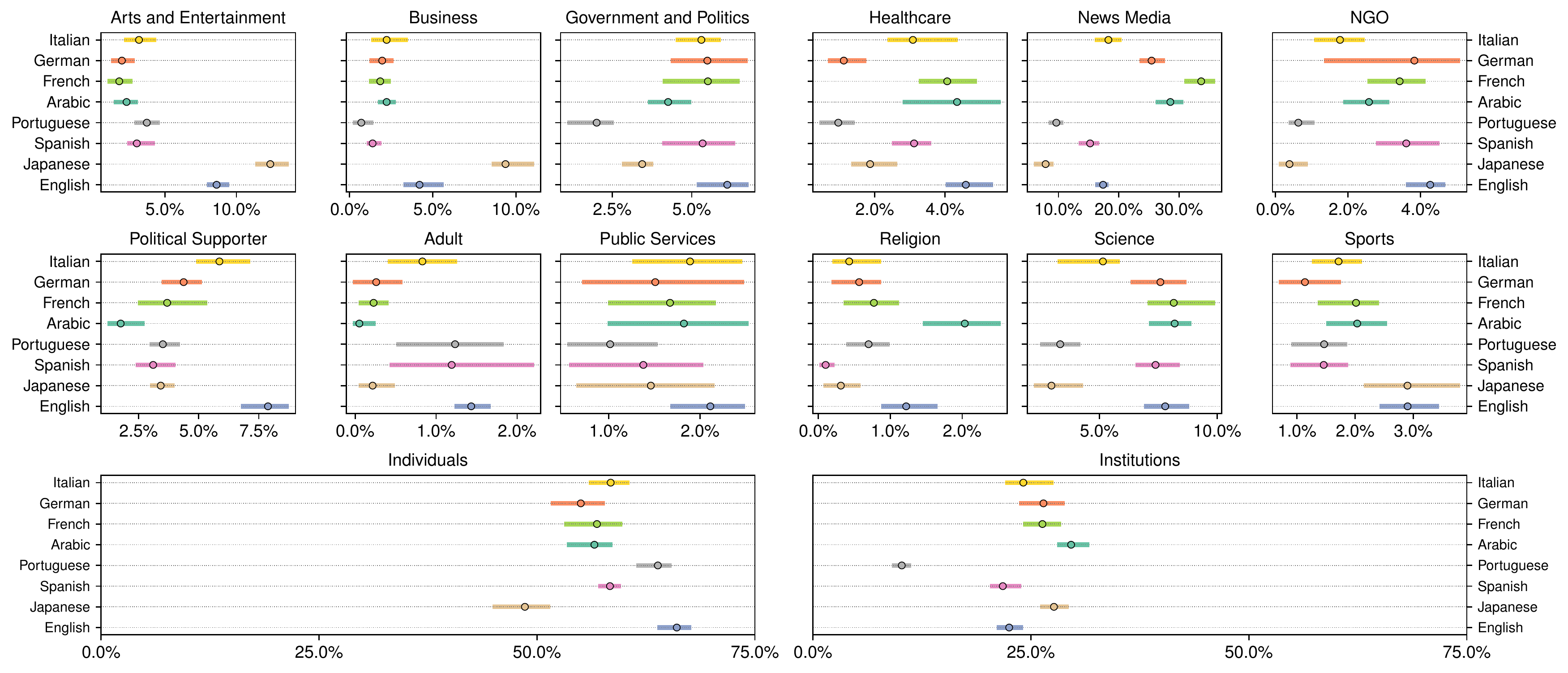}
    \caption{Category and type prevalence across different languages. }
    \label{fig:proportions}
\end{figure}

This procedure is used to obtain representative weights for Figures 1, 2, and 3 in the main text. We use the same methodology to provide supplementary view on the category and type prevalence across languages in Fig. S~\ref{fig:proportions} .

\clearpage

\section{Supplementary Tables}
\label{sec:tables}
\begin{table}[h]
\caption{The COVID-19 Twitter accounts taxonomy: \textbf{category of account}.}
\scriptsize
\begin{tabular}{ l|l} 
 \multicolumn{2}{l}{\textbf{Account category:} Please select the category that best describes this account. Use your judgement and 
  choose the one} \\
  \multicolumn{2}{l}{ that is the most suitable. In case multiple categories apply, select all that apply.}\\
 \hline
 Category of account & Description \\
 \hline
  \hline
  Media: News & Accounts related to media outlets, publishers, TV shows, radio shows, podcasts, and also \\
  & personal accounts of journalists and other communicators associated with the media outlets.  \\
  & Professionals employed by large media outlets and also accounts associated with those.\\
   \hline
  Media: Scientific  News& Accounts related to media outlets, publishers, TV shows, radio shows, podcasts, and also \\
  and Communication & personal accounts of journalists and other communicators associated with the media outlets.\\
  & Professionals employed by outlets more specific to science communication and also accounts \\
  & associated with those. \\
   \hline
  Media: Other Media & Accounts related to media outlets, publishers, TV shows, radio shows, podcasts, and also \\
  & personal accounts of journalists and other communicators associated with the media outlets.\\
  & Individuals and entities broadly related to media, but not with news. For example, podcast hosts \\
  & or fashion magazines would be in this category.\\
   \hline
  Business & Accounts associated with business such as stores, bars, restaurants, and private services like  \\
  & hair salons or gyms, and individuals associated with businesses. \\
   \hline
  Government and Politics & Accounts associated with local or national governments, political parties, and individuals who  \\
  & are closely involved with these institutions. \\
   \hline
  Public Services & Accounts associated with public services such as high schools or police departments.\\
   \hline
  NGO & Non-governmental political organization, and users who are closely involved with \\
   & these institutions. Notice that individuals in these categories are likely to be a subset of OPS, so \\
  & if there is a clear NGO that individuals support, there is no need to also label them as OPS.  \\
   \hline
 Political Supporter  & Individual accounts associated with political movements.\\
  \hline
Religion & Accounts associated with notable religious figures and religious leaders/priests as well as \\
& religious institutions and entities, tightly associated with these individuals such as temples, \\
& congregations, and online sources of religious content.\\
 \hline
Science: Engineering& Researchers, scientists, professors, graduate students, professionals, or entities representing 
\\
 and Technology& or tightly associated with these individuals. Students who are receiving education in a\\
& corresponding branch of science (except pre-med, who are categorized as healthcare). Expertise \\
& in engineering, computer science or other technology related fields.\\

 \hline
Science: Life & Researchers, scientists, professors, graduate students, professionals, or entities representing 
\\
 Sciences & or tightly associated with these individuals. Students who are receiving education in a\\
& corresponding branch of science (except pre-med, who are categorized as healthcare). Expertise \\
& in the study of biology, health and environment.\\
 \hline
Science: Social & Researchers, scientists, professors, graduate students, professionals, or entities representing 
\\
 Sciences & or tightly associated with these individuals. Students who are receiving education in a\\
& corresponding branch of science (except pre-med, who are categorized as healthcare). Expertise \\
& in the study of human societies, policies, economics.\\

 \hline
Science: Other& Researchers, scientists, professors, graduate students, professionals, or entities representing 
\\
 Sciences & or tightly associated with these individuals. Students who are receiving education in a\\
& corresponding branch of science (except pre-med, who are categorized as healthcare). Expertise \\
& in other fields. If the field of expertise is unclear, it should also be assigned to this category.\\
 \hline
Healthcare & Professionals that are employed by healthcare institutions, and that directly or indirectly take part \\
& in healthcare providing services to patients. Also includes entities representing or tightly associated \\
& with these individuals. Includes students of medicine (e.g., premed).\\
 \hline
Arts and Entertainment  & Musicians, actors, plastic artists, writers and entities representing or tightly associated with them. \\
& Notice that, similarly to sport, individuals whose hobby is art are not included. If an account belongs\\
& to an individual, art is the individual's main occupation.\\
 \hline
Sports & Athletes and entities representing or tightly associated with them, such as clubs, championships \\
& or fan accounts. E-sports are also included, so if someone is a professional video-game player, they \\
& should also be included.  If an account belongs to an individual, sport is the individual's main \\
& occupation. People whose hobby is sport are not included, e.g. having “runner” in a bio does not \\
& suggest that the individual belongs to the category.\\
 \hline
Adult Content & Accounts associated with lewd content. Producer of amateur porn, porn actors or actresses,\\
& websites related to porn, and similar.\\
 \hline
Not in English & Users whose description is not written in English. When labelling these please do not specify \\
& the type of account, that is tag them as unclear. \\
 \hline
Other & Please select this category when none of the others apply.\\
\hline
\end{tabular}

\label{tab:cat}
\end{table}

\begin{table}[h]
\small
\centering
\caption{The COVID-19 Twitter users taxonomy: \textbf{type of account}.}
\begin{tabular}{ l|l} 
 \multicolumn{2}{l}{\textbf{Account type:} Who does this account represent or belong to?} \\
 \hline
 Type of account & Description \\
 \hline
 Institution & Account clearly belongs to an institution, an official or unofficial set of individuals.  \\ 
 Individual & Account clearly belongs to an individual. \\ 
 Unclear & Account does not clearly belong to a single institution or a single individual.\\ 
 \hline
\end{tabular}
\label{tab:type}
\end{table}

\begin{table}[h]
\centering
\small
\caption{The distribution of accounts tweeting about COVID-19 in the complete one week sample, and
corresponding number of sampled and annotated accounts, across languages.}
\begin{tabular}{ l|l|l} 
 \hline
 Language & Number of unique accounts & Number of annotated accounts \\
 \hline
English & 89,652 & 1800 \\
Japanese &  33,609 & 1600 \\
Spanish &  36,033 & 1600 \\
Portuguese &  14,813 & 1500\\
Indonesian &  3291 &  1300\\
Hindi & 8165  &  1400\\
French & 4225 &  1300\\
German &  2205 & 1200\\
Italian & 1598 &  1200\\
Arabic & 3357 & 1300\\
\hline
Overall: &  196,948 & 14,200\\
 \hline
\end{tabular}
\label{tab:dist}
\end{table}

\begin{table}[h]
\centering
\small
\caption{Inter-annotator agreements.}
\begin{tabular}{ l|l|l|l|l|} 
 \hline
 Language & Category & Type \\
 \hline
 \multicolumn{2}{l}{Studies languages:} \\
 \hline
English  &  0.50  &  0.54 \\ 
Japanese  &  0.39  &  0.33 \\ 
Spanish  &  0.39  &  0.51 \\ 
Portuguese  &  0.44  &  0.30 \\
French  &  0.25  &  0.34 \\
German  &  0.34 &  0.50 \\ 
Italian  &  0.43  &  0.48 \\ 
Arabic  & 0.40 &  0.53 \\
\hline
Overall: & 0.43 & 0.44 \\
\hline
\multicolumn{2}{l}{Omitted languages:} \\
\hline
Hindi & 0.21& 0.21\\
Indonesian & 0.22& 0.24\\
\hline
\end{tabular}
\label{tab:aggr}
\end{table}

\clearpage

\section{Supplementary Figures}
\label{sec:figures}

\clearpage

\section{Supplementary Text}
\label{sec:suptext}
\begin{figure}
    \centering
    \includegraphics[width = \textwidth]{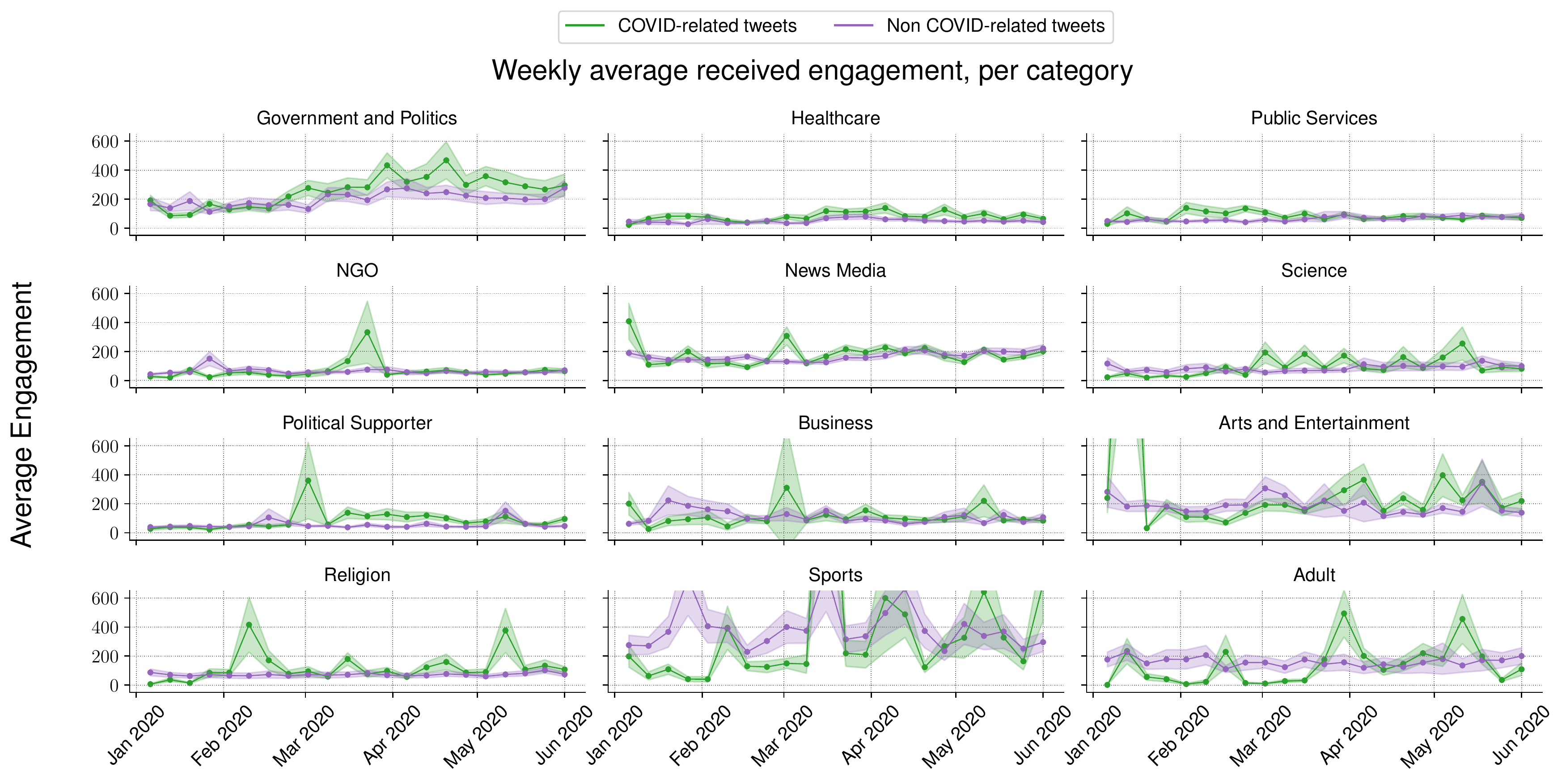}
    \caption{To further understand the mechanisms behind the change in engagement, we show the results of a complimentary analysis where we look at the effect on engagement of tweeting specifically about COVID-19, identified using the same keywords used by Twitter. COVID-19-related tweets consistently receive more engagement for Healthcare, Government and Politics and Political Supporters.}
    \label{fig:covid}
\end{figure}

\begin{figure}
    \centering
    \includegraphics[width = 0.5\textwidth]{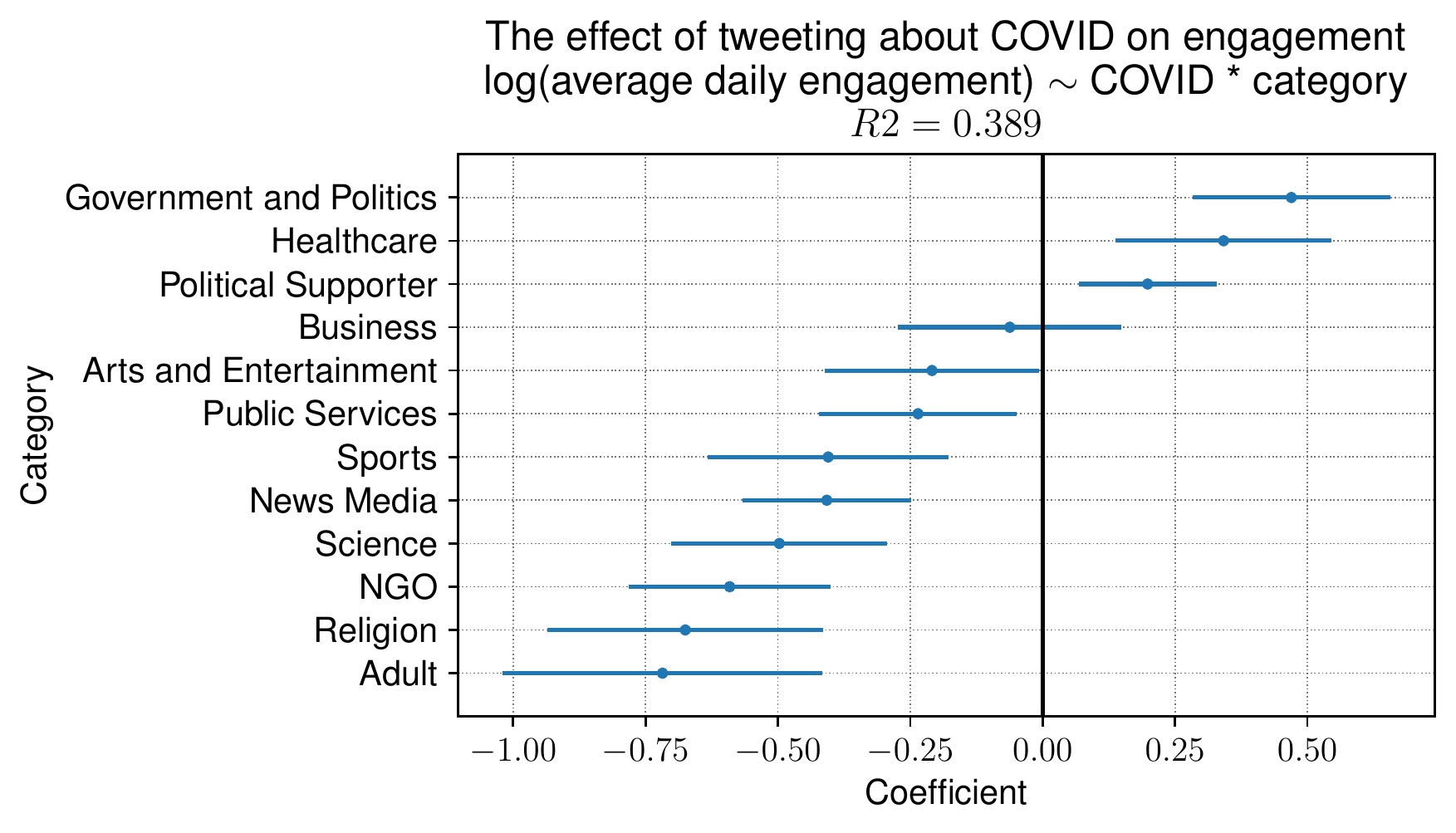}
    \caption{Daily average received engagement across categories for COVID and non-COVID tweets.}
    \label{fig:regressioncovid}
\end{figure}

\begin{figure}[h]
\centering
\begin{minipage}{0.4\textwidth}
  \centering
  \includegraphics[width=\linewidth]{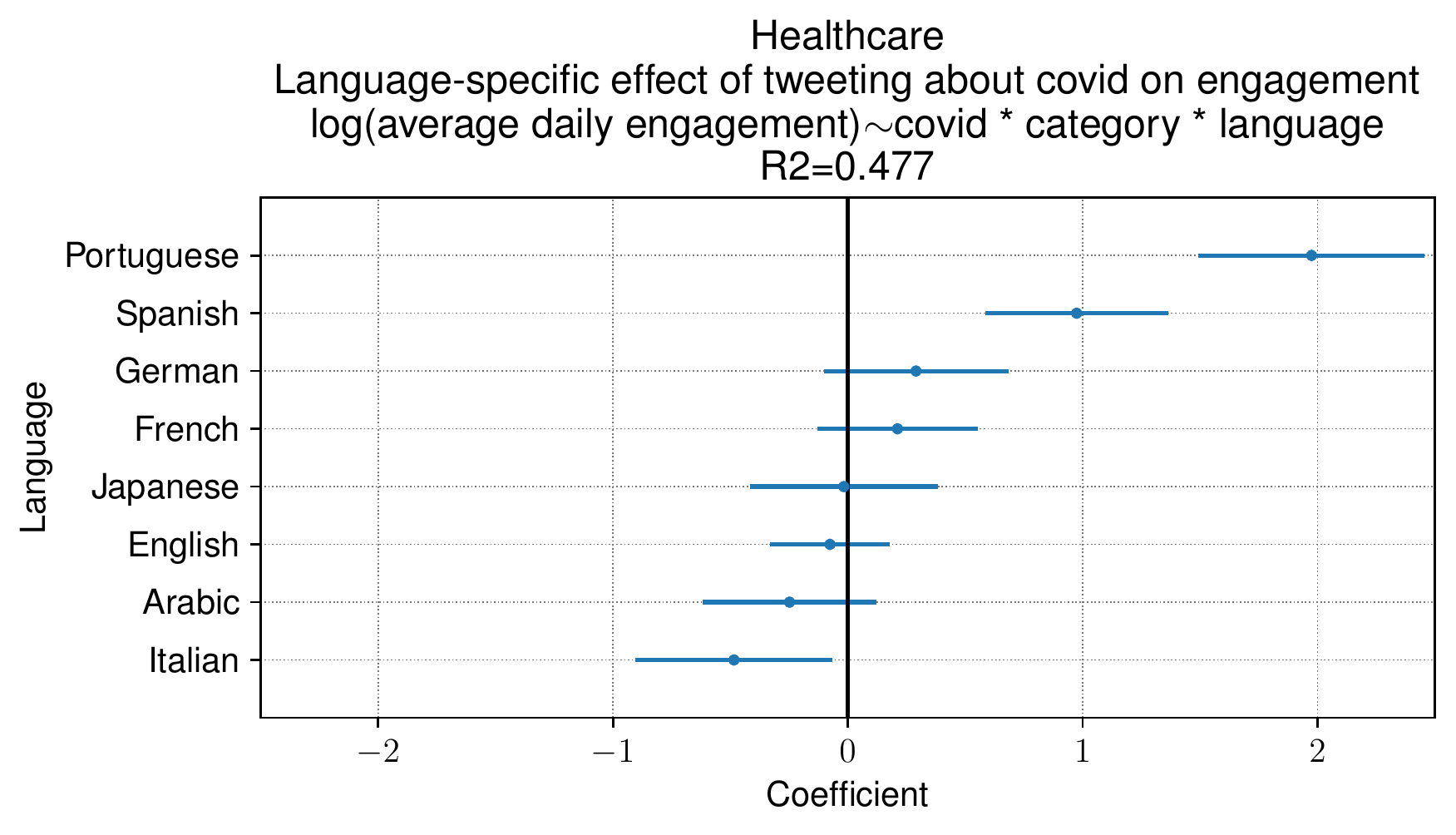}
\end{minipage}%
\begin{minipage}{0.4\textwidth}
  \centering
  \includegraphics[width=\linewidth]{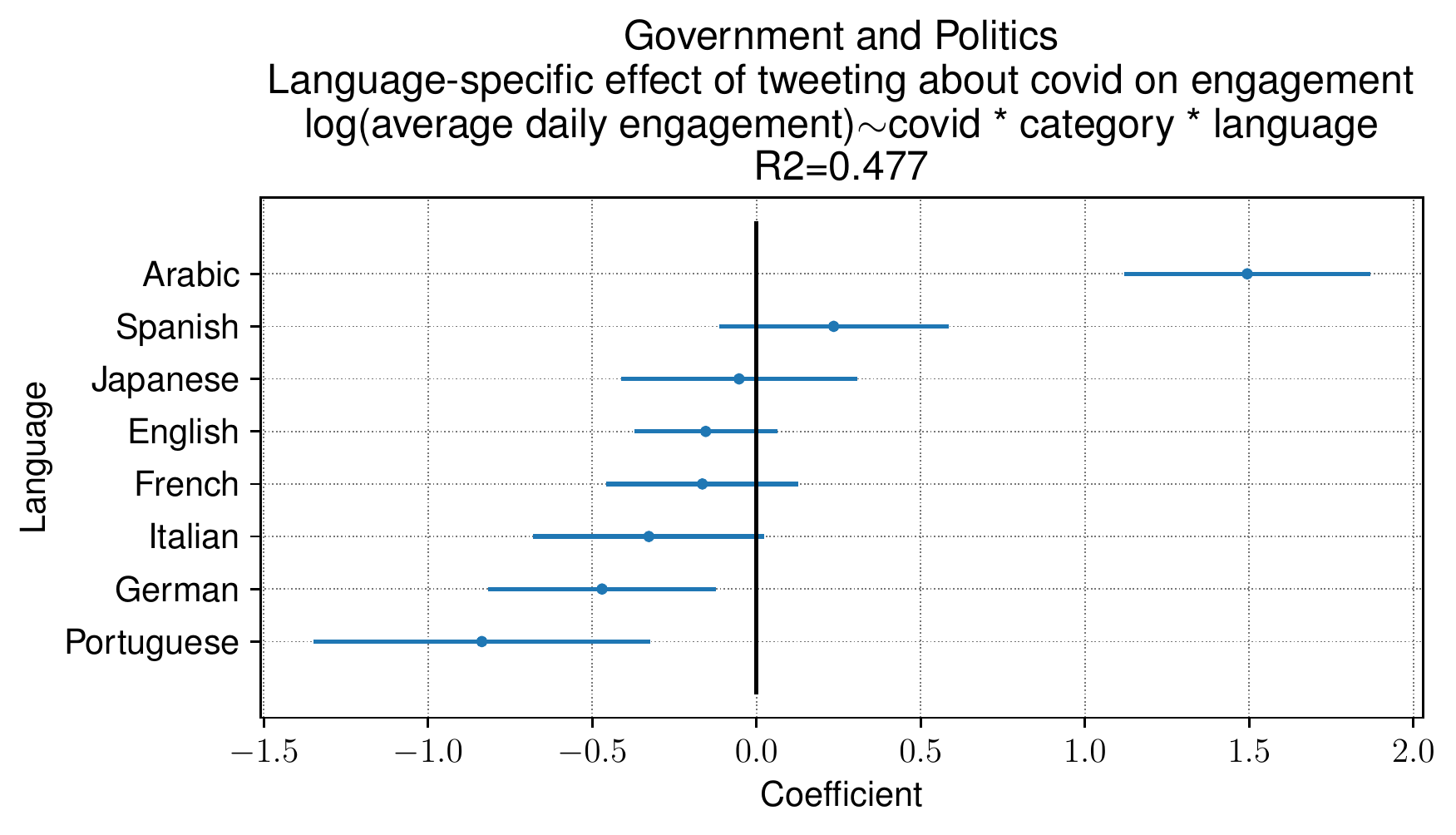}
\end{minipage}
\caption{Language-specific effect of tweeting about COVID-19 on engagement for accounts belonging to Healthcare (left), and Government and Politics (right). Important cultural differences emerge which will require future work. As an example, for COVID-19-related tweets in Portugese (largely from Brazil), Government and Politics is negatively correlated with engagement while Healthcare is positively correlated with engagement. }
\label{fig:lang}
\end{figure}

We provide a set of alternative views on the analysis discussed in Figure 2. First, in Fig. S ~\ref{fig:covid}, we show how the topic of tweets---whether they are COVID-19-related or not---plays a role in determining the degree of engagement they received during the pandemic, with COVID-19-related tweets consistently receiving more engagement for Healthcare, Government \& Politics and Political Supporters. We support this analysis with regression modelling, presented in Fig. S ~\ref{fig:regressioncovid} and Fig. S ~\ref{fig:lang}.

\begin{figure}
    \centering
    \includegraphics[width = \textwidth]{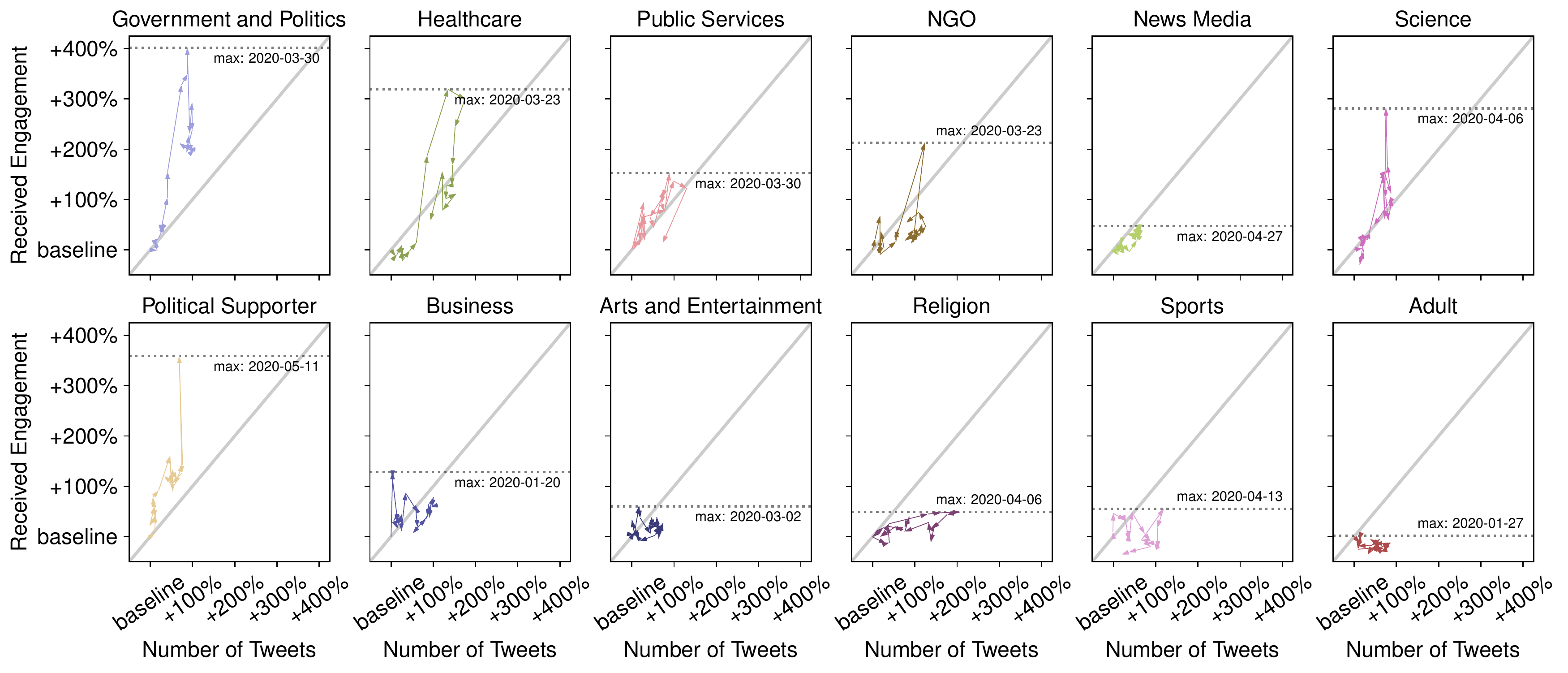}
    \caption{We show an alternate view of the analysis depicted in Figure 2 of the main text. Here,each week of the Study Period is sequentially connected by arrows in a 2D-plane where the x-axis depicts the weekly average increase in volume, and the y-axis the weekly average increase in engagement.
}
    \label{fig:paths}
\end{figure}

\begin{figure}[t]
    \centering
    \begin{subfigure}
    \centering
    \includegraphics[width=.45\linewidth]{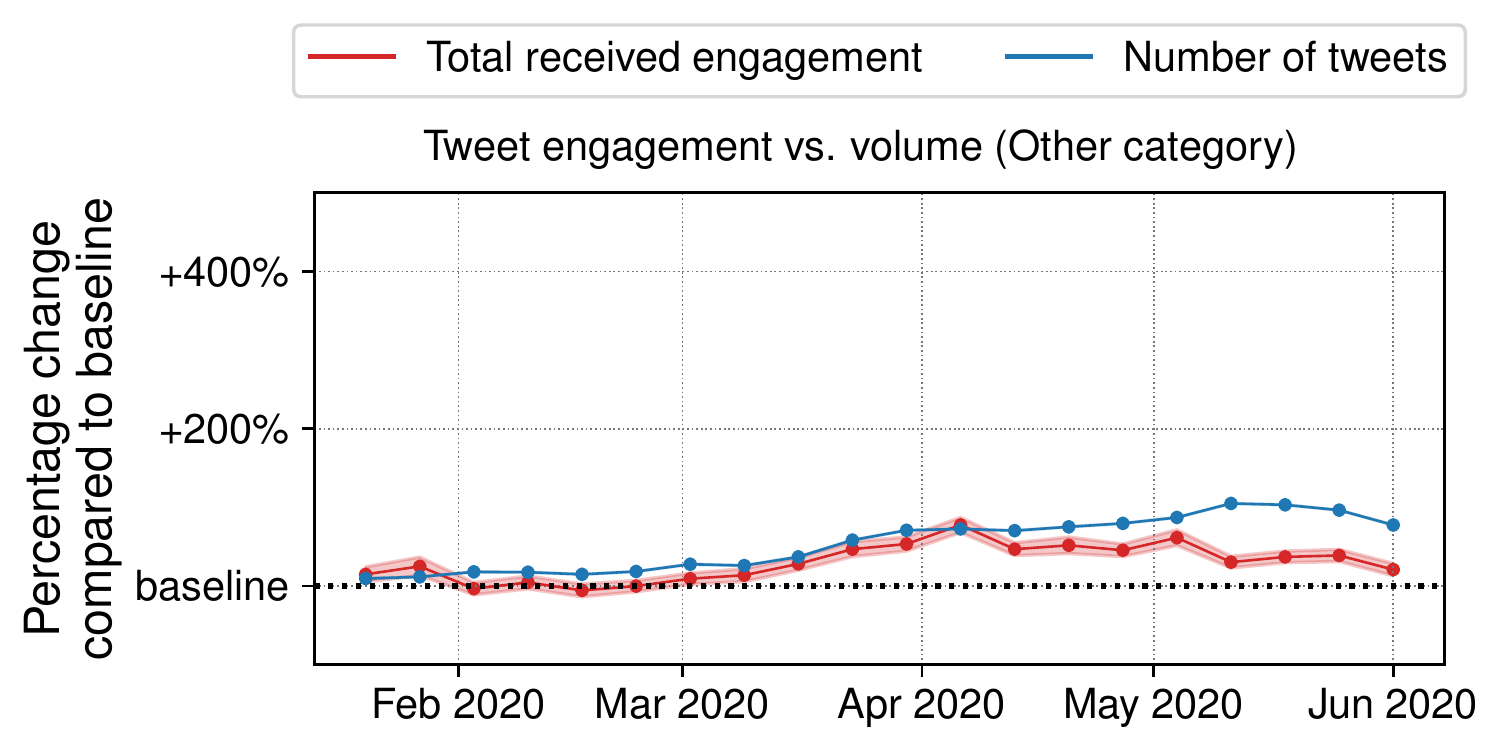}
        \caption*{\small \textbf{(a)} The account-averaged percentage change (calculated with IPW) in number of tweets (in blue) and engagement (in red). The change is shown relative to a baseline, calculated using the two weeks of January 2020.}
        \label{fig:a}
    \end{subfigure}%
    \begin{subfigure}
    \centering
    \includegraphics[width=.45\linewidth]{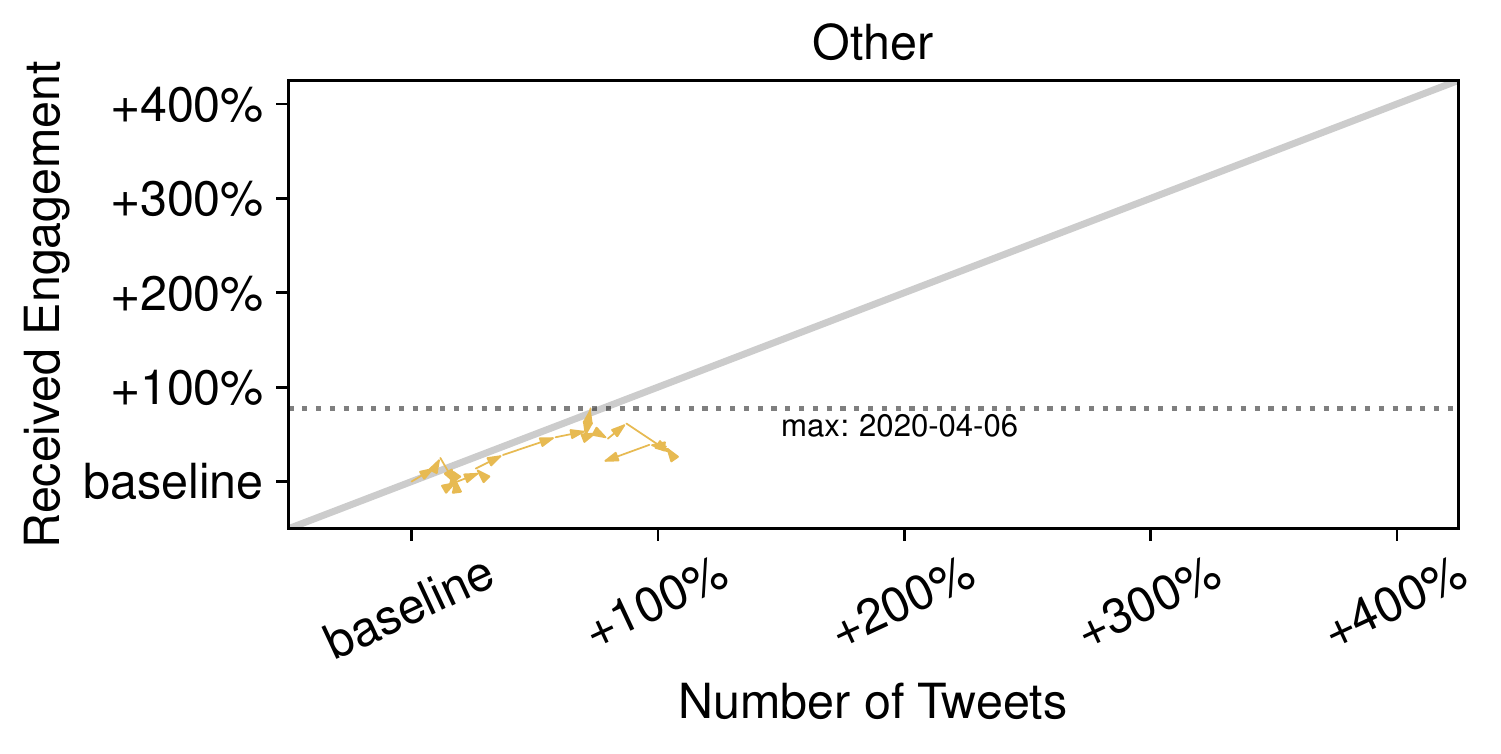}
        \caption*{\small \textbf{(b)} Each week of the Study Period is sequentially connected by arrows in a 2D-plane where the x-axis depicts the weekly average increase in volume, and the y-axis the weekly average increase in engagement.}
        \label{fig:a}
    \end{subfigure}%
    \begin{subfigure}
    \centering
    \includegraphics[width=.45\linewidth]{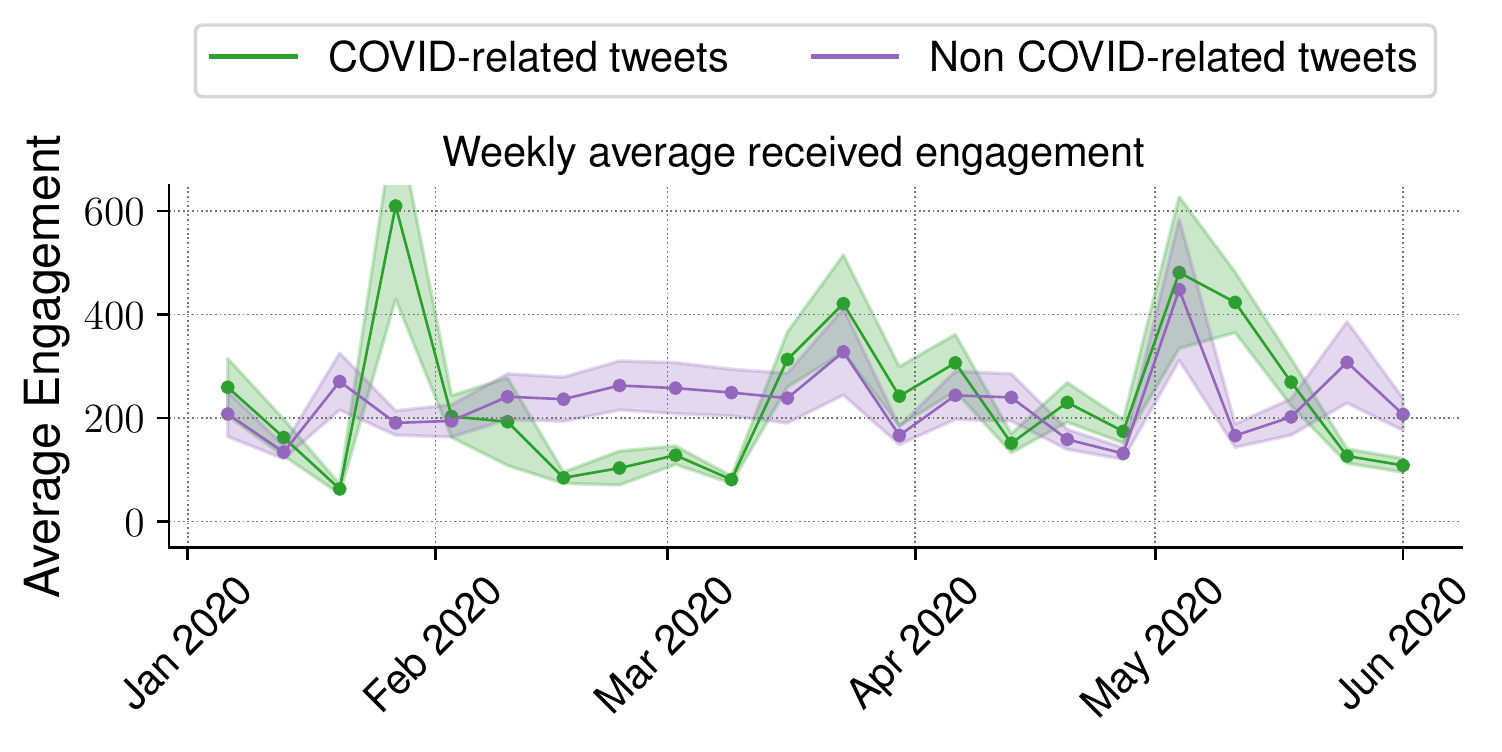}
        \caption*{\small \textbf{(c)} The effect on average engagement of tweeting specifically about COVID-19.}
        \label{fig:c}
    \end{subfigure}
    \caption{{Figures 2, S\ref{fig:paths} and S\ref{fig:covid} for category Other, a category with unremarkable differences compared to categories with pronounced surges in engagement.}}
    \label{fig:other}
\end{figure}

\begin{figure}
    \centering
    \includegraphics[width = \textwidth]{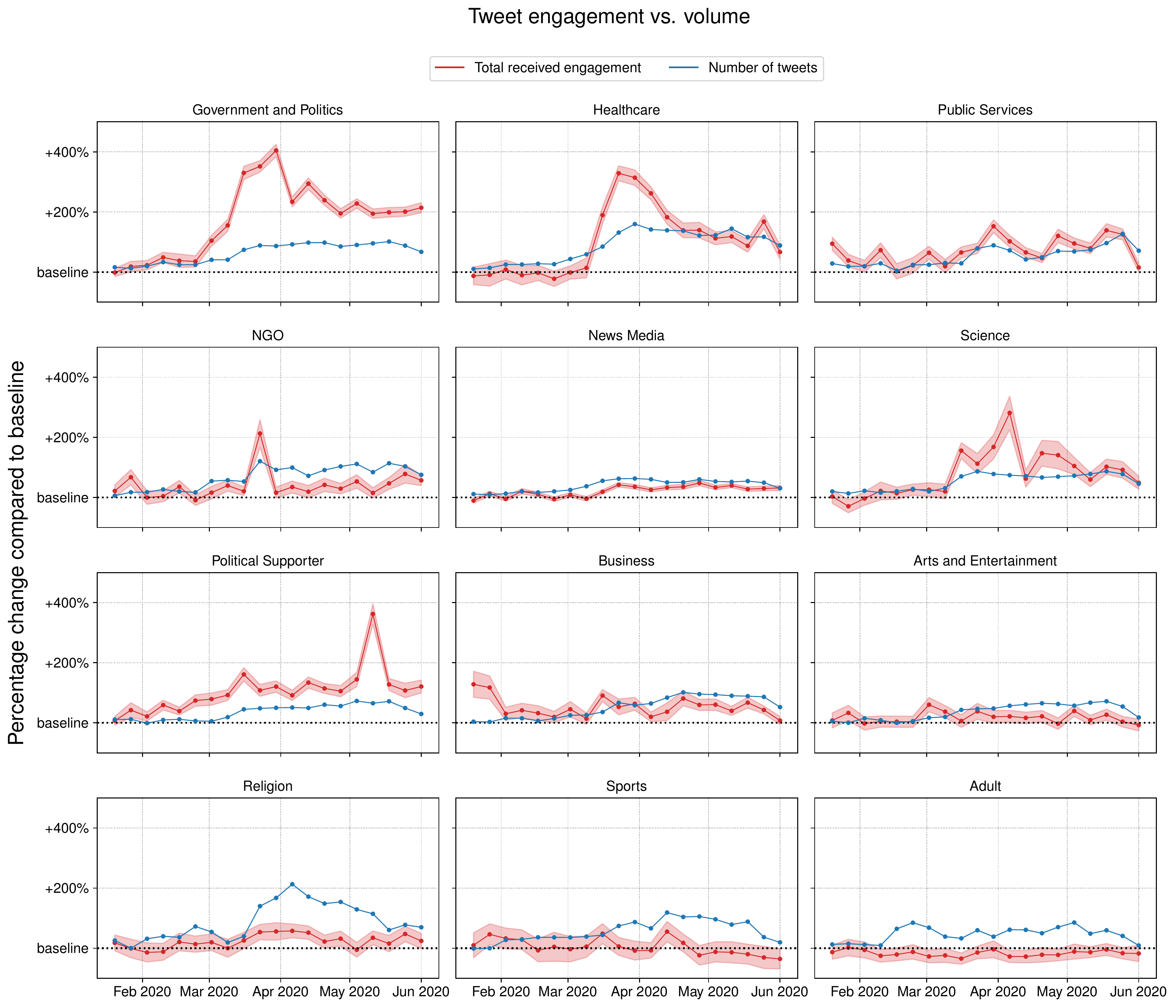}
    \caption{To alleviate a potential bias in our analysis caused by the hypothetical surge of new users joining the platform during the crisis, we conducted an alternative analysis where we restricted ourselves to a set of users who created their Twitter account before the studied period. In that way, we excluded 482, out of 14000 annotated users. The observed trends are not impacted by the presence of such newcomers.}
    \label{fig:onlyold}
\end{figure}

In Fig. S ~\ref{fig:paths}, we provide an alternative view of Fig. 2, where each week of the Study Period corresponds to a point connected by an arrow with the previous week. In Fig. S ~\ref{fig:other}, we provide the results for the Other category, which is excluded in the analysis. Similarly, in Fig. ~\ref{fig:onlyold}, we examine the robustness of our findings by evaluating the impact of users joining the platform during the Studied Period.

Lastly, we provide additional information about a set of supplemental experiments, with the goal of understanding, first, the degree of automated activity within the studied accounts, and second, the between-category interactions that drive the trends in engagement. In order to measure which categories retweet which other categories, we use an automated method for label expansion, detailed below.

\subsection{Bot detection}
To assess the degree of bot activity in our data, we used the tool Botometer\footnote{\scriptsize\url{https://botometer.iuni.iu.edu}}.
Botometer uses a supervised Machine Learning approach to estimate the so called complete automation probability (CAP), for which a value of 1 indicates complete automation.
Botometer extracts features from recent tweets in the account's timeline, such as temporal activity patterns, social networks and sentiment, among others.
In this work, we use a CAP threshold of 0.25 in order to decide whether a account is presumed to be a bot.
The bot activity data was collected via the Botometer API between July 22 and July 27, 2020.
By using the method above on a sample of 5000 accounts in our annotation dataset (dataset A), we find around 3.3\% of presumed automated accounts.
Bot activity in the annotation dataset was significantly higher for accounts annotated as "other" (4.6\% bots) for the category labels and "unclear" (5.3\%) in the type of account labeling (that is, when annotators had to classify accounts as belonging to an individual or an institution).
Testing was performed using a one-tailed binomial test at significance level $\alpha=0.5$ (before Bonferroni correction).
Based on these numbers, bots seem to only have a marginal influence on the overall validity of the results which are based on the sampled user accounts (dataset B).

\subsection{Who retweets whom?}

In Fig. S~\ref{fig:whortwho}, we looked at all tweets and retweets produced in the week of interest. Recall that here we take advantage of the fact that the data obtained from the stream is complete, that is we are certain to have all the retweets of a given tweet.
We deploy an automatic classifier described better in Section 1.3, to automatically label the category of all accounts in the week of interest.
Excluded are accounts with user descriptions of less than 3 characters, yielding labels for a total of 39.2M users.
With the labels generated by the classifier, we build a retweet digraph $G$. 
Each node $u$ in this graph is an account, assigned to a single category (the most likely according to the classifier).
Each $(u,v)$ edge in this graph stands for a retweet from account $u$ to account $v$.  
That means that an edge only exists if the tweet by account $v$ was retweeted more than 10 times.

Given this graph, we proceeded to explore the number of retweets between categories. 
This can be thought of as a collapsed graph $G'$ where all nodes with the same category are collapsed into one. 
Looking at this graph we analyze, for each category, where are the sources of the incoming edges. 

We also obtain a null model with this graph. 
This null model assumes that each category is equally likely to connect to any other category.
Thus, suppose we want to calculate the percentage of incoming edges from category $X$ to category $Y$.
Let $Out(X)$ be the number of outgoing edges from category $X$ and $In(Y)$ be the number of incoming edges from category $Y$. 
Also, let $B$ be the total number of edges in the graph. Notice that $In(Y)/B$ is the fraction of all edges that are incoming edges towards $Y$. If the assignment of edges from category $X$ is really independent of other categories, we would expect $Out(X)  \frac{In(Y)}{B}$ edges between $X$ and $Y$. 
This what we consider to be our category-agnostic random null model.

Lastly, to obtain confidence intervals over this analysis we bootstrap the whole process, we choose a random sample of the edges in the original graph $G$ to "collapse" generating the category-graph $G'$.
We repeat this procedure 1000 times, and obtain confidence intervals for the expected value and the observed value for each category.

\subsection{Label Expansion}
\begin{figure}
    \centering
    \includegraphics[width = \textwidth]{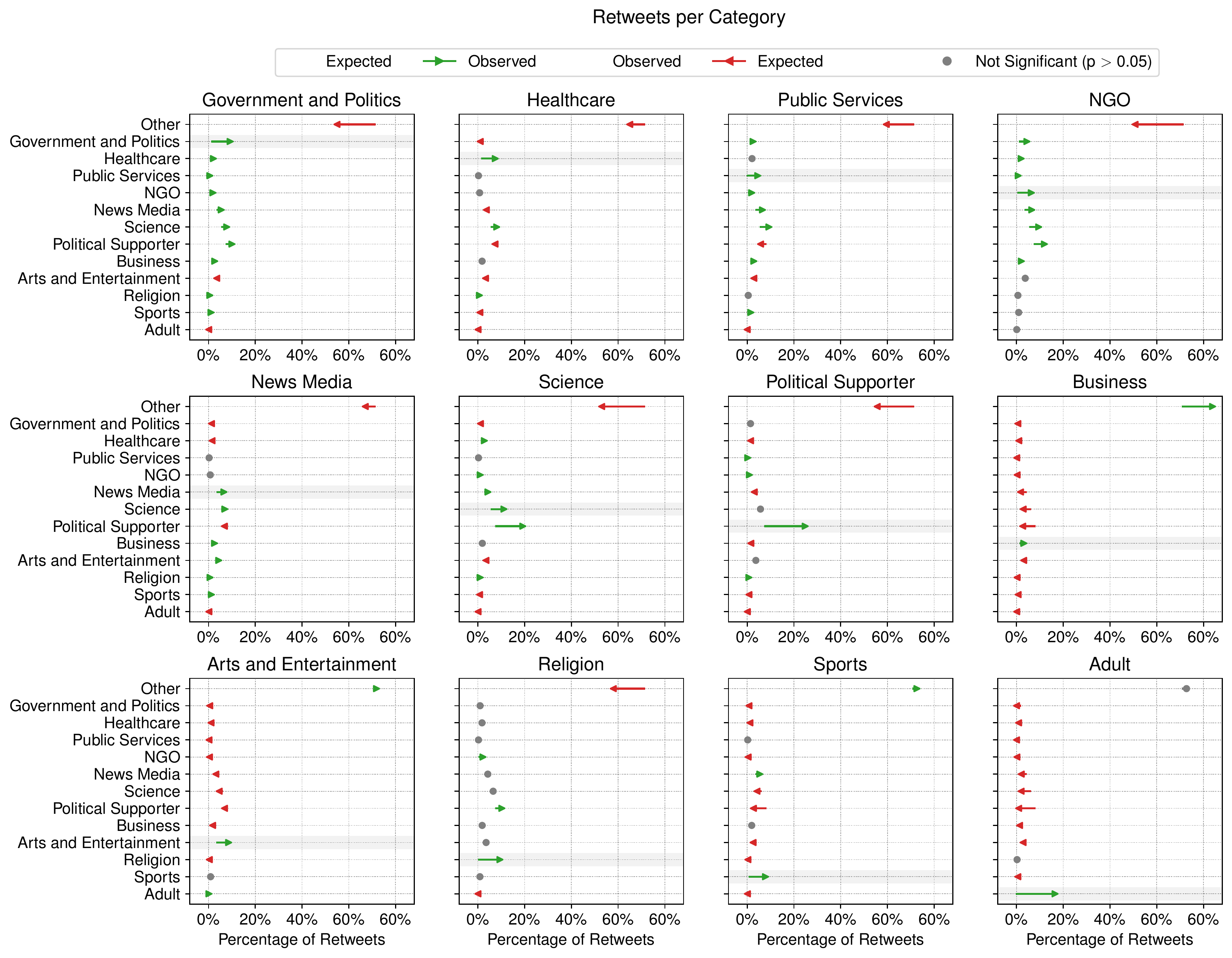}
    \caption{We measure, for each category, what is the source of their engagement (who retweets them). We compare this value to a category agnostic null model that assumes each category receives engagement at random, proportionally to their size. The figure shows arrows that start from the expected value, according to the null model, and end at the observed value. Where differences are not significant (p>0.05), arrow ends are replaced by gray circles. The Other category, while being numerically larger, is a net retweeter of the remaining categories and gets retweeted less frequently than expected. We also observe a strong homophily: all categories retweet significantly (p < 0.05) more tweets from their own category than predicted by the null model. The one exception are Political Supporters retweeting Science more than Science retweeting itself.}
    \label{fig:whortwho}
\end{figure}

In Fig. S~\ref{fig:whortwho}, we used label expansion, a method in which a Machine Learning classifier is trained on the subset of annotated data to predict the labels for the full data set.
The account descriptions consist of unstructured text, including frequent use of emojis, and special Unicode characters.
Furthermore, the entire COVID-19 Twitter stream data is multilingual, covering 41 languages from very diverse language families.
Given this complexity, two major approaches were tried using the FastText library\footnote{\scriptsize{\url{https://github.com/facebookresearch/fastText}}} and models based on the BERT family (Bidirectional Encoder Representations from Transformers)~\footnote{\scriptsize{\url{https://github.com/google-research/bert}}}.

\subsubsection{BERT}
BERT is a general-purpose language understanding model which can be used, among other applications, for text classification. 
BERT models are pretrained on large bodies of plain text (e.g. from Wikipedia) in an unsupervised way. 
Pretrained models can then be used in a supervised downstream task, such as text classification, in a process called finetuning.

In this work, we started with the pretrained multilingual cased BERT model (\texttt{bert-multilang}), a BERT model which was simultaneously pretrained on the Wikipedia corpora of 104 languages.
However, our target domain (Twitter account descriptions) is very different from text found in Wikipedia.
Therefore, an additional step of unsupervised pretraining, also called domain-specific pretraining, was conducted on our existing corpus of account descriptions.
account descriptions from dataset A of at least 3 characters length, which were not contained in the annotation dataset (dataset B), were combined into a dataset of 52M account descriptions, comprising of a total of 697M tokens.
The data was preprocessed by replacing account names, URLs, and email addresses with generic fillers.
Furthermore, emojis were replaced by textual versions (e.g. the American flag emoji would be replaced by \texttt{:flag-us:}), using the Python emoji library\footnote{\scriptsize \url{https://pypi.org/project/emoji/}}.
From this dataset 593M training examples were generated. 
Training was run for roughly 1 epoch (600k steps) at a batch size of 1024 and a constant learning rate of 2e-5.
Training took roughly 2 days on a TPU v3-8 (8 cores, 128 GB of memory), and resulted in a new model, which we refer to as \texttt{bert-multilang-pt}.

A similar procedure was applied for a English-only model, in which pretraining was conducted with account descriptions in English (251M training examples, 21M account descriptions), and pretraining was started from the English BERT-large uncased (whole word masking) variant (\texttt{bert-english}).
Training for this variant was conducted with the same batch size and learning rate, but for roughly 2 epochs (roughly 5 days of training).
We will refer to this variant as \texttt{bert-english-pt}.

\subsubsection{FastText}
FastText is a lightweight library for text classification and representation learning.
It is a shallow model that uses subword information to enrich word vectors.
Similar to BERT, it is possible to fine-tune pretrained word representations for text classification purposes.
In contrast to BERT, which heavily relies on training on GPUs, it can be trained on a large dataset using multicore CPUs in a matter of minutes.
Also, FastText models are much more compact than BERT (in our case, 125 MB vs 700 MB).

For FastText models, we only used account descriptions in English language.
Preprocessing was conducted by normalizing texts, replacing accoun tnames, URLs and emails and removing emojis.
We then pretrained a FastText skipgram model for 5 epochs, with a learning rate of 0.1, context window size of 5, and n-gram size between 3 and 6.
We will refer to the pretrained FastText model as \texttt{fasttext-english-pt} .

\subsubsection{Finetuning}
Eventually all pretrained models were finetuned on the type (3 classes) and category (13 classes) tasks.
The annotation data was deduplicated (accounts may have identical descriptions), and preprocessed in the same way the the pretraining data was prepared for the respective model type.
The preprocessed annotation data (100\%, $n_\text{category}=9913$, $n_{\text{type}}=10725$) was split into a training (64\%), development (16\%), and test set (20\%) for both type and category, respectively. 
Multilingual models were fine-tuned on the original training data, whereas English models were fine-tuned on the translated versions of the account descriptions.
Model selection was performed by optimizing the respective F1-macro score on the development set.

BERT-like models were fine-tuned in 10 epochs, using a learning rate of 1e-5 (using 10\% warm-up with linear decay) and training batch sizes of 32. 

FastText models were fine-tuned using built-in hyperparameter autotuning available for supervised training with a vector dimension of 100.

\subsubsection{Classifier results}
Based on the pretrained models described above, we compare downstream classifier performance scores in Fig. S~\ref{fig:compare_runs}.  
Unexpectedly, BERT models trained on English-only data outperform the multilingual BERT model.
Generally, we also see a performance boost due to domain-specific pretraining.
The best English-only model (\texttt{bert-english-pt}) gives a F1-macro score of 0.71 and 0.62, on the category and type datasets, respectively.
The smaller FastText models (\texttt{fasttext-english-pt}) perform comparably to other models on the type dataset but give slightly lower scores on the category dataset.
The best multilingual model (\texttt{bert-multilang-pt}) yields F1-macro scores of 0.56 (category) and 0.63 (type).

\begin{figure}
    \centering
    \begin{subfigure}
    \centering
    \includegraphics[width=.45\linewidth]{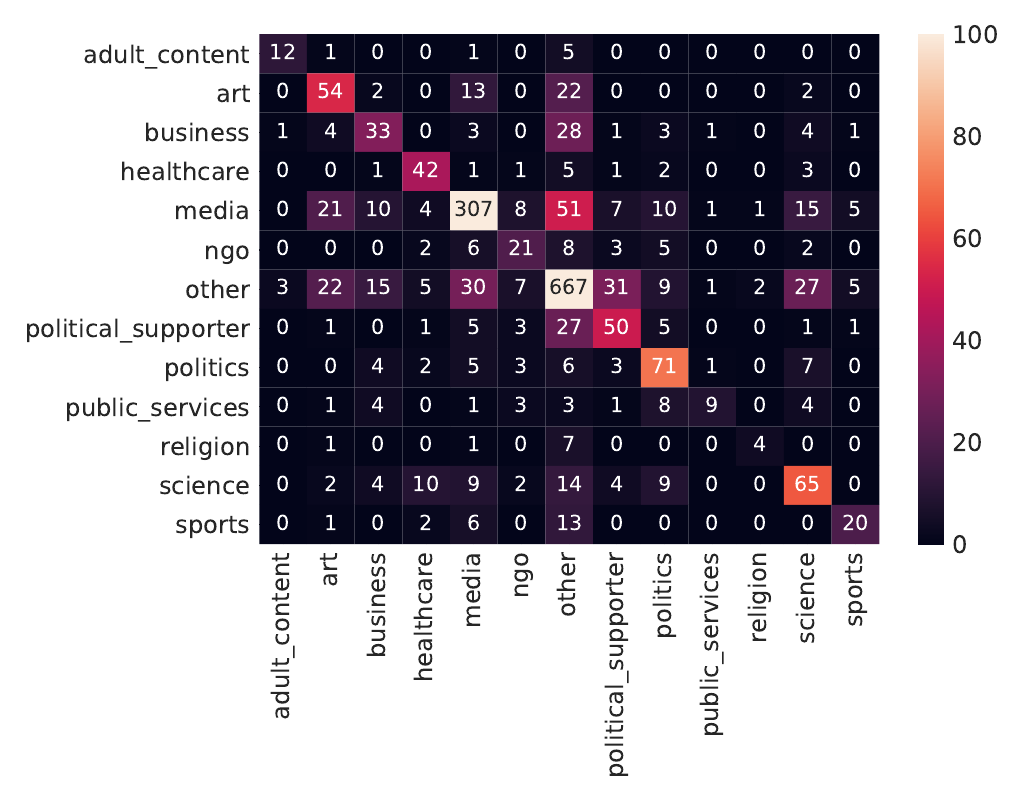}
    \label{fig:sub1}
    \end{subfigure}%
    \begin{subfigure}
    \centering
    \includegraphics[width=.45\linewidth]{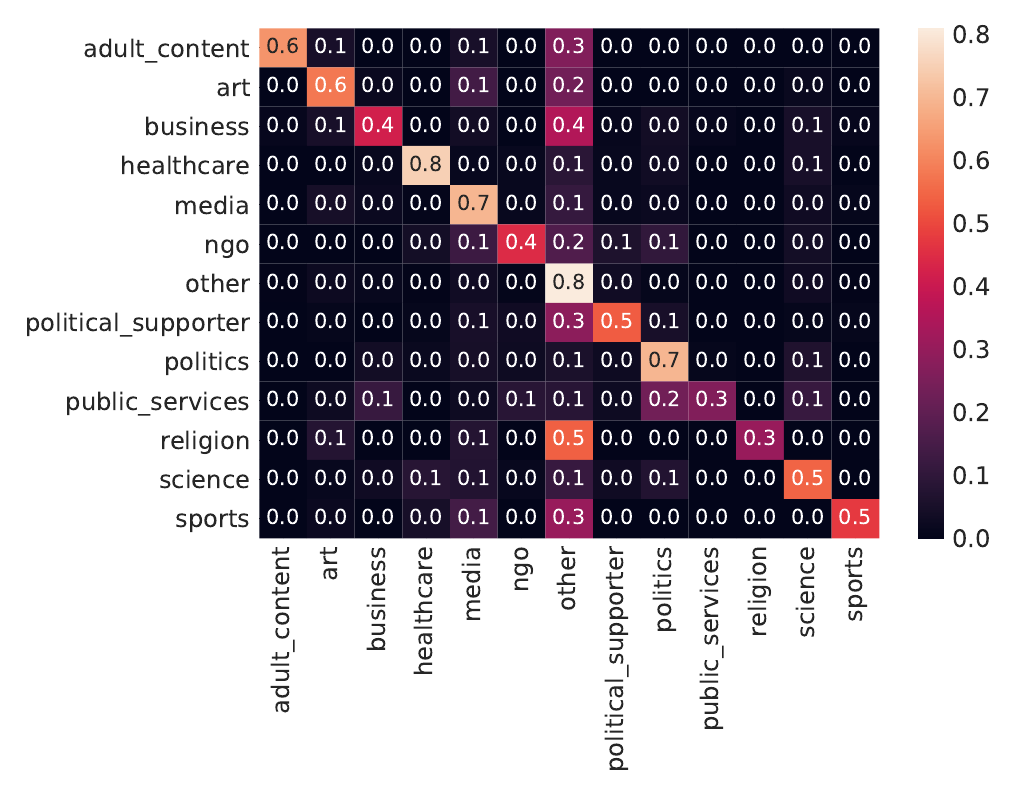}
    \label{fig:sub2}
    \end{subfigure}
    \caption{
        Confusion matrix on the held out test set for the multilingual BERT category classifier (fine-tuned version of \texttt{bert-multilang-pt}).
        The y-axis represents the true label (as per annotation data) and the x-axis represents the label predicted by the classifier.
        Confusion matrix on the left shows absolute counts, whereas on the right normalized counts are shown.
        Most errors were made by predicting a account description as "other" (which was the most frequent category).
        The weakest categories are "religion" (often predicted as "other"), and "public\_services" (often predicted as "politics").
        These categories also have had few training and test examples.
    }
    \label{fig:confusion_category}
\end{figure}

\begin{figure}
    \centering
    \begin{subfigure}
    \centering
    \includegraphics[width=.4\linewidth]{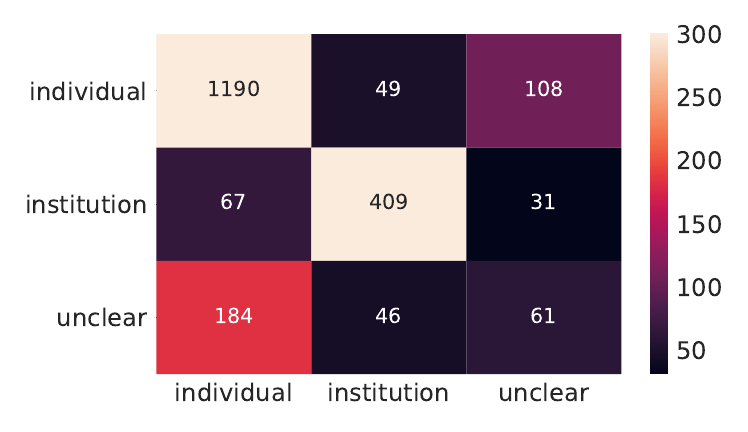}
    \end{subfigure}%
    \begin{subfigure}
    \centering
    \includegraphics[width=.4\linewidth]{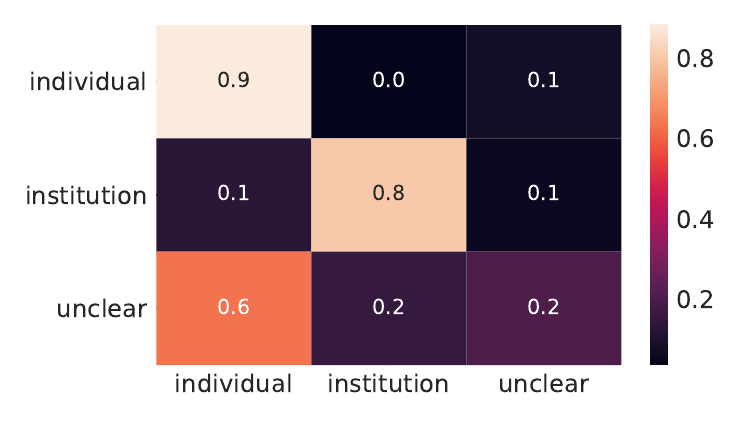}
    \end{subfigure}
    \caption{
        Confusion matrix on the held out test set for the BERT type classifier (fine-tuned version of \texttt{bert-multilang-pt}).
        The y-axis represents the true label (as per annotation data) and the x-axis represents the label predicted by the classifier.
        Confusion matrix on the left shows absolute counts, whereas on the left normalized counts are shown.
        Predictions for "individual" and "institution" are very accurate.
        "Unclear" represents a relatively small class, therefore leading to a higher relative error.
    }
    \label{fig:confusion_type}
\end{figure}

\begin{figure}
    \centering
    \begin{subfigure}
    \centering
    \includegraphics[width=.45\linewidth]{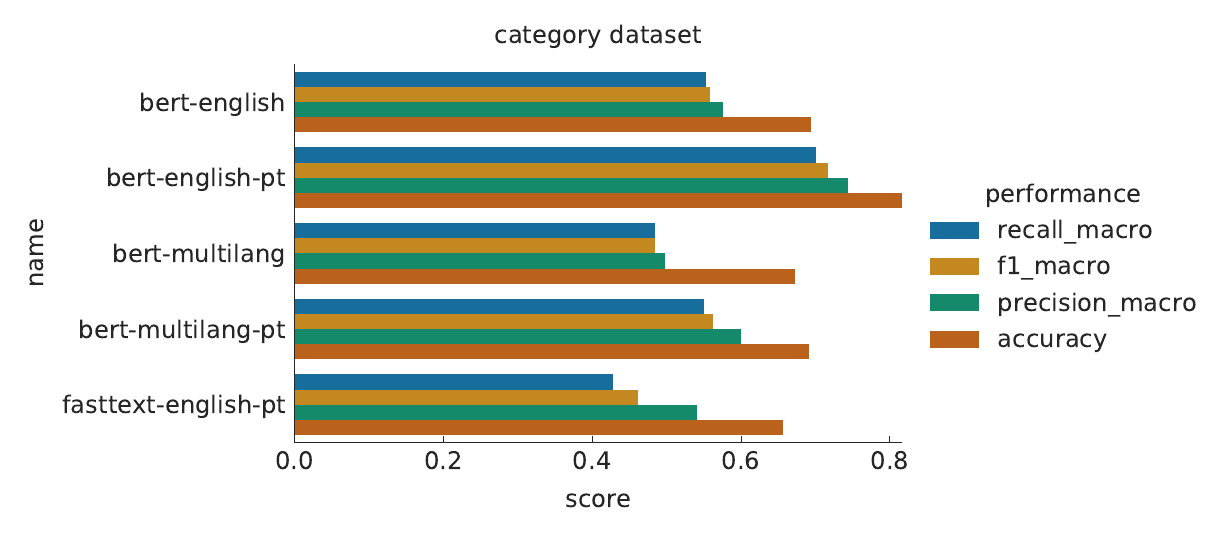}
    \end{subfigure}%
    \begin{subfigure}
    \centering
    \includegraphics[width=.45\linewidth]{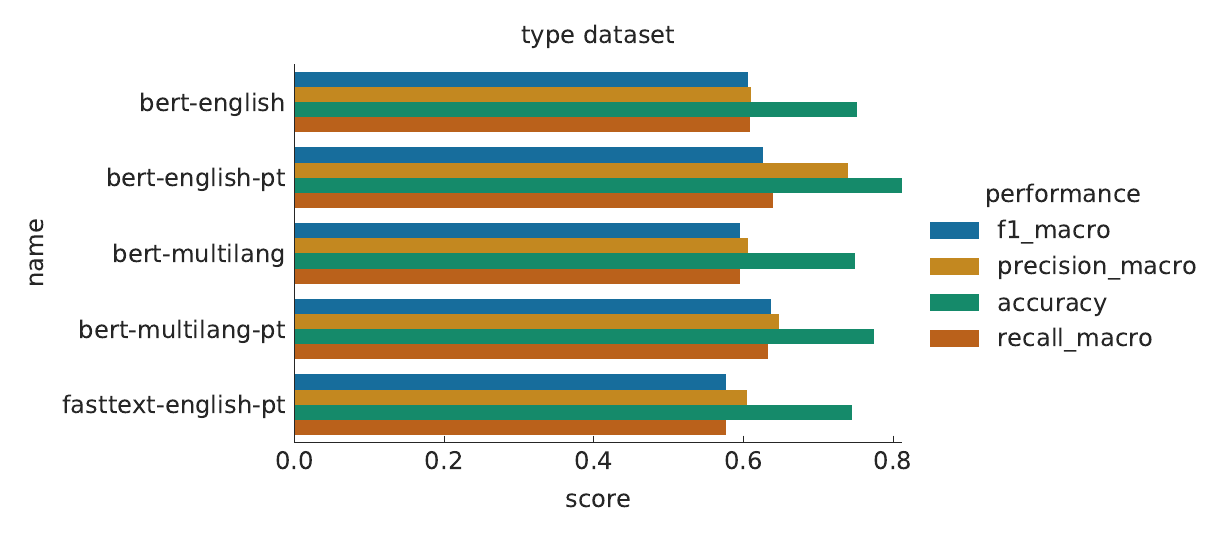}
    \end{subfigure}
    \caption{
        Comparisons of test scores of BERT and FastText classifiers.
        Overall, best results are achieved for English-only models.
        Models which underwent domain-specific pretraining, as indicated by the "pt" suffix, generally outperform the default pretrained models.
        The model used for the analysis is \texttt{bert-multilang-pt}.
    }
    \label{fig:compare_runs}
\end{figure}

For further analysis we focus on the multilingual BERT model (\texttt{bert-multilang-pt}), which was the final model used for label expansion in this work. 
When inspecting the confusion matrices (Fig. S~\ref{fig:confusion_category} and Fig. S~\ref{fig:confusion_type}), classifier scores for this model are generally satisfying.
Certain classes for which only very few observations are present show lower scores in comparison.
In particular, this is concerning the classes "Religion" and "Public Services" (for category) and "Unclear" (for type). 
The smallest error rates can be expected for the classes "Healthcare", "News Media", and "Government and Politics".
No significant deviations from the mean accuracy could be observed for individual languages.
Testing was performed using a two-sided binomial test at significance level $\alpha=0.5$ (before Bonferroni correction).

\clearpage

\end{document}